\RequirePackage{fix-cm} 
\documentclass[a4paper, twoside, 12pt, dvips, reqno]{amsart}

\usepackage{fixltx2e}     

\usepackage{amssymb}
\usepackage{amsmath}

\usepackage{a4wide}

\usepackage{indentfirst}
\usepackage{graphicx}
\usepackage{psfrag}

\usepackage[usenames,dvipsnames]{pstricks}
\usepackage{epsfig}
\usepackage{pst-grad} 
\usepackage{pst-plot} 

\usepackage{subfigure}

\usepackage{longtable}

\usepackage[english]{babel}
\usepackage[latin1]{inputenc}

\usepackage{color}
\definecolor{oneblue}{rgb}{0,0.0,0.75}
\usepackage[colorlinks,
            urlcolor=oneblue,
            linkcolor=oneblue,
            citecolor=oneblue,
            bookmarksopen=false,
            pagebackref]{hyperref}

\numberwithin{equation}{section}

\newtheorem{remark}{Remark}

\newcommand{\x}{\vec{x}}
	\newcommand{\n}{\hat{n}}
\renewcommand{\k}{\vec{k}}
\renewcommand{\u}{\vec{u}}
\newcommand{\grad}{\nabla}
\newcommand{\phis}{\varphi}
\newcommand{\R}{\mathbb{R}}
\newcommand{\D}{\mathcal{D}}
\newcommand{\Hv}{\mathcal{H}}
\newcommand{\A}{\mathcal{A}}
\newcommand{\B}{\mathcal{B}}
\renewcommand{\P}{\mathbb{P}}
\renewcommand{\L}{\mathcal{L}}
\renewcommand{\O}{\mathcal{O}}
\newcommand{\N}{\mathcal{N}}
\newcommand{\F}{\mathcal{F}}
\newcommand{\dz}{\partial_z}
\newcommand{\dt}{\partial_t}
\newcommand{\Fr}{\mathrm{Fr}}
\newcommand{\phih}{\hat{\phi}}
\renewcommand{\div}{\nabla\cdot}
\newcommand{\od}[2]{\frac{d #1}{d #2}}
\newcommand{\half}{{\textstyle{1\over2}}}
\newcommand{\sech}{\mathop{\mathrm{sech}}}
\newcommand{\pd}[2]{\frac{\partial#1}{\partial#2}}

\begin{document}

\title[Finite fault solution for tsunami generation]{On the use of the finite fault solution for tsunami generation problems}

\author[D. Dutykh]{Denys Dutykh$^*$}
\address{LAMA, UMR 5127 CNRS, Universit\'e de Savoie, Campus Scientifique,
73376 Le Bourget-du-Lac Cedex, France}
\email{Denys.Dutykh@univ-savoie.fr}
\urladdr{http://www.lama.univ-savoie.fr/~dutykh/}
\thanks{$^*$ Corresponding author}

\author[D. Mitsotakis]{Dimitrios Mitsotakis}
\address{IMA, University of Minnesota, Minneapolis, MN 55455, USA}
\urladdr{http://sites.google.com/site/dmitsot/}

\author[X. Gardeil]{Xavier Gardeil}
\address{LAMA, UMR 5127 CNRS, Universit\'e de Savoie, Campus Scientifique,
73376 Le Bourget-du-Lac Cedex, France}
\email{Xavier.Gardeil@etu.univ-savoie.fr}

\author[F. Dias]{Fr\'ed\'eric Dias}
\address{School of Mathematical Sciences, University College Dublin, Belfield, Dublin 4, Ireland}
\email{frederic.dias@ucd.ie}

\begin{abstract}
The present study is devoted to the problem of tsunami wave generation. The main goal of this work is two-fold. First of all, we propose a simple and computationally inexpensive model for the description of the sea bed displacement during an underwater earthquake, based on the finite fault solution for the slip distribution under some assumptions on the dynamics of the rupturing process. Once the bottom motion is reconstructed, we study waves induced on the free surface of the ocean. For this purpose we consider three different models approximating the Euler equations of the water wave theory. Namely, we use the linearized Euler equations (we are in fact solving the Cauchy-Poisson problem), a Boussinesq system and a novel weakly nonlinear model. An intercomparison of these approaches is performed. The developments of the present study are illustrated on the 17 July 2006 Java event, where an underwater earthquake of magnitude 7.7 generated a tsunami that inundated the southern coast of Java.
\end{abstract}


\maketitle

\section{Introduction}

Tsunami waves have attracted a lot of attention by researchers. The interest of the scientific community has especially increased after the two megatsunamis in December 2004 \cite{Syno2006}, where nearly 230,000 people in fourteen countries lost their lives, and in March 2011, where 20,000 people lost their lives in Japan. The 2004 event also led Indian Ocean countries to develop Tsunami Warning Systems (TWS) \cite{Synolakis2005, Basher2006}, unfortunately more on an individual basis than on a collective basis. The most elaborated warning system to date is the Pacific Ocean TWS, which has been developed over several decades by efforts of NOAA's specialists \cite{Titov2005, Gonz}. 

An operational tsunami wave modeling tool is an essential part of any warning system \cite{Titov2005, Tkalich2007a}. Mathematical and numerical models in use should be constantly improved to produce more accurate results in less CPU time \cite{Imamura1996, Titov1997, Dutykh2009a}. In order to study the propagation of a tsunami wave, an initial condition must usually be provided to any numerical model designed for this purpose. The present study is an attempt to improve the construction of the initial tsunami waveform. The set of existing practices described in the literature constitutes the field of the so-called tsunami generation modeling \cite{Hammack, Todo, Dutykh2006, Dutykh2007a, Dutykh2007b, Dutykh2008}.

The modeling of tsunami generation was initiated in the early sixties by the prominent work of Kajiura \cite{kajiura}, who proposed the translation of the static sea bed displacement towards the free surface as an initial condition. Classically, the celebrated Okada \cite{Okada85, okada92} and sometimes Mansinha \& Smylie\footnote{In fact, the Mansinha \& Smylie solution is a particular case of the more general Okada solution.} \cite{Mansinha1967, Mansinha1971} solutions are used to compute the co-seismic sea bed displacements. This approach is still widely used by the tsunami wave modeling community. However, significant progress has been recently made in this direction \cite{Ohmachi2001, Dutykh2006, Dutykh2007a, Dutykh2007b, Rabinovich2008, Saito2009, Dutykh2009a}.

In the present study we exploit some recent advances in seismology to reconstruct better co-seismic displacements of a tsunamigenic earthquake. More precisely, we suggest using the so-called finite fault solution developed by Ji and his collaborators \cite{Bassin2000, Ji2002}, based on static and seismic data inversion. This solution provides multiple fault segments of variable local slip, rake angle and several other parameters. By applying Okada's solution to each subfault, we reconstruct the sea bed displacement with higher resolution. To our knowledge, this technique has already been employed to model the Kuril islands tsunamis of 15 November 2006 and 13 January 2007, cf. \cite{Rabinovich2008}. Since Okada's solution consists of relatively simple closed-form analytical expressions, all computations can be done efficiently enough so that they can be used in a real-time TWS (cf. \cite{Weinstein2008}). The obvious \textit{sine qua non} condition is that the corresponding finite fault inversion should also be performed in a reasonable time.

In the present study we go further in reconstructing the dynamic sea bed displacement according to the rupture propagation speed and the rise time also provided by the finite fault solution. Constructed in this special way, sea bed displacements are then coupled with several water wave models. Among them, there is a novel weakly nonlinear solver based on a formulation involving the Dirichlet-to-Neumann operator which is computed approximately using Fourier transforms. The other two models considered here are the linearized free surface Euler equations and a Boussinesq type system. Developments presented in this paper are illustrated on the example of July 17, 2006 Java event \cite{Ammon2006}. However, we would like to stress that the methodology presented in this study is quite general and can be applied to many other tsunamigenic earthquakes for which a finite fault solution is available.

The paper is organized as follows. In Section \ref{sec:displ} we describe the static and dynamic sea bed displacements, while in Section \ref{sec:fluid} we present a simple approximate water wave solver with a moving bottom. In Section \ref{sec:numres} we study numerically the generation process of a real-world event. An intercomparison of the three models mentioned above is performed. Some important conclusions are drawn in Section \ref{sec:concl}.

\section{Co-seismic displacement construction}\label{sec:displ}

\begin{figure}
  \centering
  \includegraphics[width=0.9\textwidth]{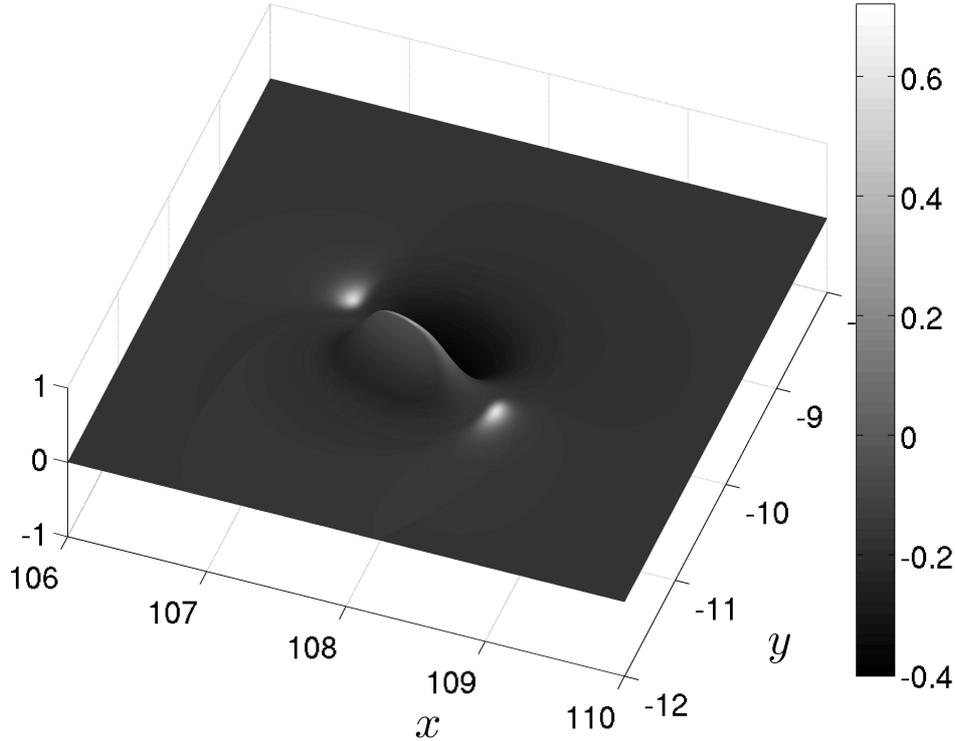}
  \caption{Static vertical displacement in meters of the seabed computed with the single fault parameters provided in Table \ref{tab:singleF}. The maximum lift is 0.7215 m while the maximum subsidence is 0.4030 m. The $x-$axis is the longitude while the $y-$axis is the latitude. The $y-$axis points to the North.}
  \label{fig:Java1_fault}
\end{figure}

The modeling of tsunami generation is directly related to the problem of the bottom motion during an underwater earthquake. Traditionally, Okada's solution \cite{Okada85, okada92} is used in regimes characterized by an active fault of small or intermediate size, i.e. consisting of one or a few segments (e.g. the great Sumatra 2004 earthquake, \cite{Syno2006, Ioualalen2007}). In this case the resulting vertical displacement field is translated to the free surface. This approach is conventionally referred to as {\em passive tsunami generation} \cite{ddk}, contrary to the {\em active generation} which explicitly involves the bottom motion dynamics \cite{Dutykh2006}. Since our methods will be illustrated on the example of the July 17, 2006 Java  event, we show in Figure \ref{fig:Java1_fault} a typical single-fault based initial condition used for the corresponding tsunami wave modeling \cite{Yalciner2008}. The seismic parameters used to produce this vertical displacement are given in Table \ref{tab:singleF}.

\begin{table}
  \begin{center}
   \begin{tabular}{c||c}
     Fault length, km & $80.9$ \\
     \hline
     Fault width, km & $40.0$ \\
     \hline
     Focal depth, km & $20.0$ \\
     \hline
     Slip, m & $2.5$ \\
     \hline
     Dip angle & $10^\circ$ \\
     \hline
     Slip angle & $95^\circ$ \\
     \hline
     Strike angle (clockwise from N) & $289^\circ$ \\
   \end{tabular}
   \caption{Seismic fault parameters for the Java 2006 event. The corresponding seismic moment can be taken as $M_0 = 2.52\times 10^{27}$ N$\cdot$ m ($M_w = 7.56$).}
   \label{tab:singleF}
  \end{center}
\end{table}

\begin{remark}
The celebrated Okada solution \cite{Okada85, okada92} is based on two main ingredients --- the dislocation theory of Volterra \cite{volt} and Mindlin's fundamental solution for an elastic half-space \cite{mindl1}. Particular cases of this solution were known before Okada's work, for example the well-known Mansinha \& Smylie's solution \cite{Mansinha1967, Mansinha1971}. Usually, all these particular cases differ by the choice of the dislocation and Burger's vector orientation \cite{press}. We recall the basic assumptions behind this solution:
\begin{itemize}
  \item The fault is immersed into a linear homogeneous and isotropic half-space
  \item The fault is a Volterra type dislocation
  \item The dislocation has a rectangular shape
\end{itemize}
For more information on Okada's solution we refer to \cite{Dutykh2006, Dias2006, Dutykh2007a}. 
\end{remark}

The finite fault solution is based on the multi-fault representation of the rupture \cite{Bassin2000, Ji2002}. The rupture complexity is reconstructed using a joint inversion of the static and seismic data. The fault's surface is parametrized by multiple segments with variable local slip, rake angle, rise time and rupture velocity. The inversion is performed in an appropriate wavelet transform space. The objective function is a weighted sum of $L_1$, $L_2$ norms and some correlative functions. With this approach seismologists are able to recover rupture slip details \cite{Bassin2000, Ji2002}. This available seismic information is exploited in this study to compute the sea bed displacements produced by an underwater earthquake with higher geophysical resolution.

\begin{figure}
\centering
\subfigure[Top view.]{\includegraphics[scale=0.35]{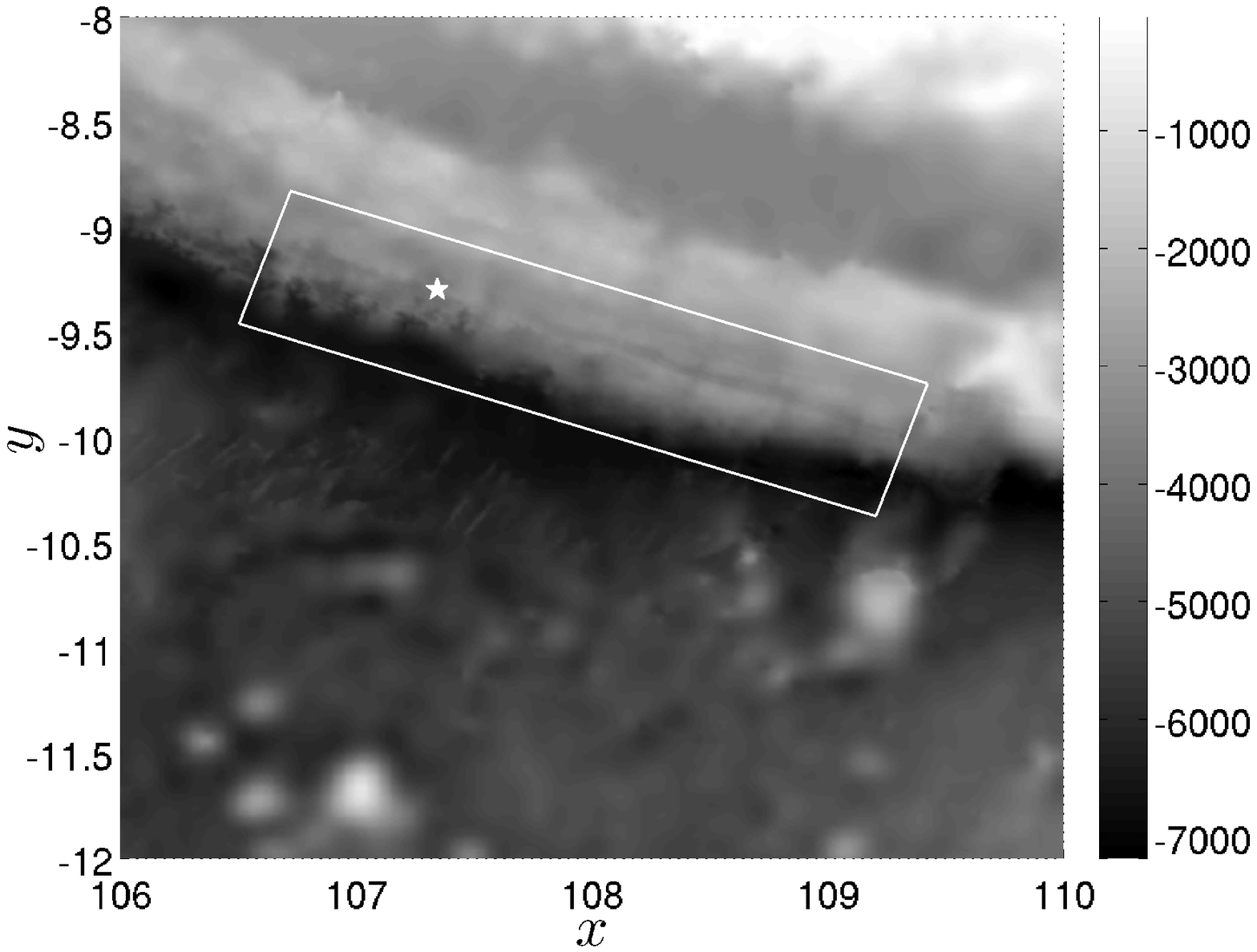}}
\subfigure[Bathymetry side view.]{\includegraphics[scale=0.35]{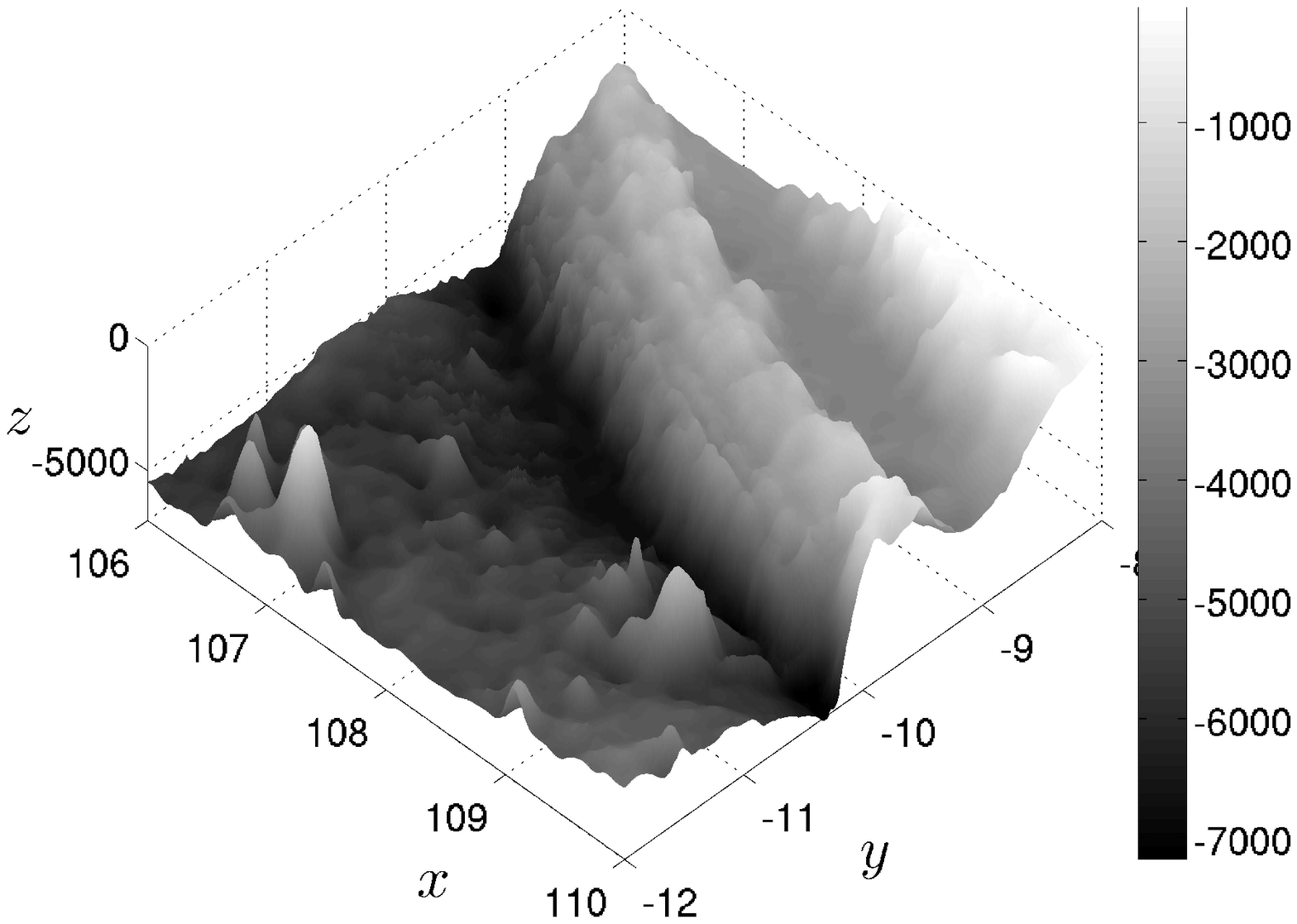}}
  \caption{Surface projection of the fault's plane and the ETOPO1 bathymetric map of the region under investigation. The symbol $\star$ indicates the epicenter's location at $(107.345^\circ, -9.295^\circ)$. The local Cartesian coordinate system is centered at the point $(108^\circ, -10^\circ)$. The region is located between $(106^\circ, -8^\circ)$ and $(110^\circ, -12^\circ)$.}
  \label{fig:Java_fault}
\end{figure}

The proposed approach will be directly illustrated on the Java 2006 event. The July 17, 2006 Java earthquake involved thrust faulting in the Java trench and generated a tsunami wave that inundated the southern coast of Java \cite{Ammon2006, Fritz2007}. The estimates of the size of the earthquake (cf. \cite{Ammon2006}) indicate a seismic moment of $6.7 \times 10^{20}$ N$\cdot$ m, which corresponds to the magnitude $M_w = 7.8$. Later this estimate was refined to $M_w = 7.7$ \cite{Ji2006}. Like other events in this region, this 2006 event had an unusually low rupture speed of $1.0$ -- $1.5$ km/s, and occurred near the up-dip edge of the subduction zone thrust fault. According to Ammon {\em et al}, most aftershocks involved normal faulting \cite{Ammon2006}. The rupture propagated approximately $200$ km along the trench with an overall duration of approximately $185$ s. The fault's surface projection along with ocean ETOPO1 bathymetric map are shown in Figure \ref{fig:Java_fault}. We note that the Indian Ocean bathymetry considered in this study varies between 7186 and 20 meters in the shallowest region.

\begin{remark}
The estimate of the finite fault inversion for this earthquake was also performed by the Caltech team \cite{Ozgun2006}. The magnitude estimated in that study was $M_w = 7.9$. In this study we do not present numerical simulations using their data but it is straightforward to apply our algorithms to this case as well.
\end{remark}

\subsection{Static displacement}

In order to illustrate the advantages of the proposed approach we will also compute the static co-seismic displacements using the finite fault solution \cite{Ji2006}. The fault is considered to be the rectangle with vertices located at $(109.20508^\circ$ (Lon), $-10.37387^\circ$ (Lat), $6.24795$ km (Depth)$)$, ($106.50434^\circ$, $-9.45925^\circ$, $6.24795$ km), ($106.72382^\circ$, $-8.82807^\circ$, $19.79951$ km), ($109.42455^\circ$, $-9.74269^\circ$, $19.79951$ km) (see Figure \ref{fig:Java_fault}a). The fault's plane is conventionally divided into $N_x = 21$ subfaults along strike and $N_y = 7$ subfaults down the dip angle, leading to a total number of $N_x\times N_y = 147$ equal segments. Parameters such as subfault location $(x_c, y_c)$, depth $d_i$, slip $u$ and rake angle $\phi$ for each segment are given in Table \ref{tab:subfaults} and can also be downloaded from \cite{Ji2006}. The elastic constants common to all subfaults and parameters such as dip and slip angles are given in Table \ref{tab:crust}. (We note that the slip angle is measured conventionally in the counter-clockwise direction from the North. The relations between the elastic wave celerities $c_p$, $c_s$ and the Lam\'e coefficients $\lambda$, $\mu$ used in Okada's solution are given in Appendix \ref{app:iii}.)

\begin{table}
\begin{center}
\begin{tabular}{c||c}
  $P$-wave celerity $c_p$, m/s & 6000 \\
  \hline
  $S$-wave celerity $c_s$, m/s & 3400 \\
  \hline
  Crust density $\rho$, kg/m$^3$ & 2700 \\
  \hline
  Dip angle, $\delta$ & $10.35^\circ$ \\
  \hline
  Strike angle (clockwise from N) & $289^\circ$ \\
\end{tabular}
\caption{Geophysical parameters used to model elastic properties of the subduction zone in the region of Java.}
\label{tab:crust}
\end{center}
\end{table}

\begin{figure}
  \centering
  \includegraphics[width=0.9\textwidth]{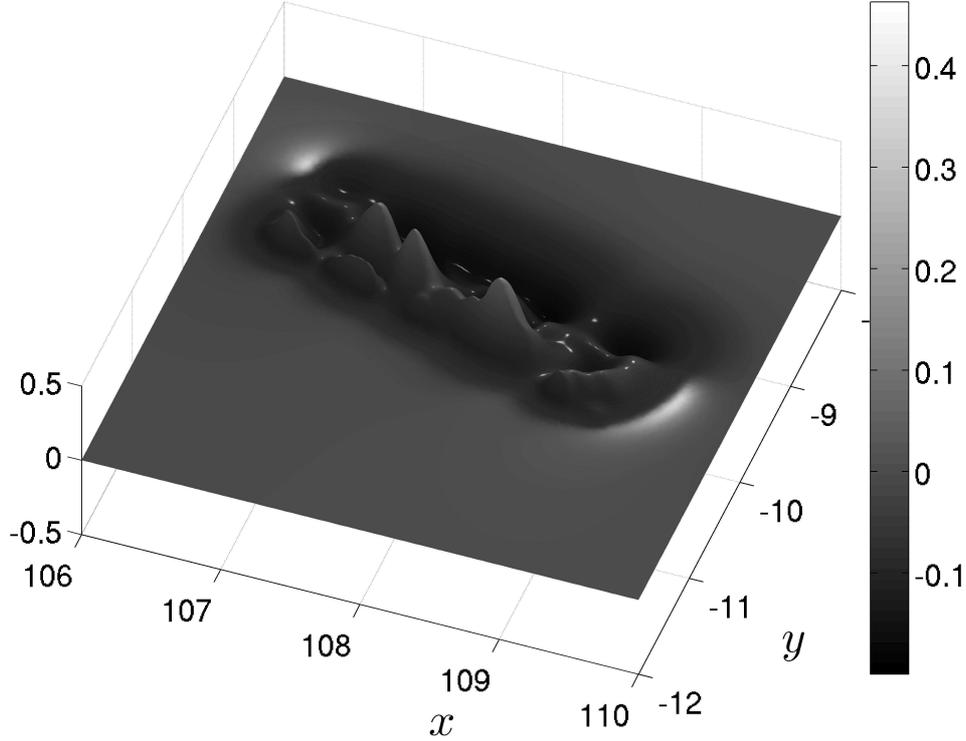}
  \caption{The vertical displacement of the finite fault solution, cf. \cite{Ji2006}. The corresponding seismic moment is  $M_0 = 3.53\times 10^{27}$ N$\cdot$ m ($M_w = 7.65$). The maximum lift is 0.4629 while the maximum subsidence is 0.1997.}
  \label{fig:ffaultdisp}
\end{figure}

We compute Okada's solution at the sea bottom by substituting $z=0$ in the geophysical coordinate system and taking the vertical component of the displacement field $\O_i(\x; \delta, \lambda, \mu, \ldots)$, where $\delta$ is the dip angle, $\lambda$, $\mu$ are the Lam\'e coefficients (see Appendix \ref{app:iii}) and the dots denote the dependence of the function $\O(\x)$ on eight other parameters, cf. \cite{Dutykh2006}. The resulting co-seismic vertical bottom displacement $\zeta(\x)$ can be computed as a simple superposition of subfault contributions:
\begin{equation*}
  \zeta (\x) = \sum_{i=1}^{N_x\times N_y} \O_i(\x; \delta, \lambda, \mu, \ldots)
\end{equation*}
The graph of $\zeta (\x)$ is presented in Figure \ref{fig:ffaultdisp}. The specific static displacement can be compared with the single fault classical approach depicted on Figure \ref{fig:Java1_fault}. It is worth mentioning that more than one local extrema can be found in this solution due to a higher slip resolution.

Hereafter we will adopt the short-hand notation $\O_i(\x)$ for the vertical displacement component of Okada's solution for the $i^{\mathrm{th}}$ segment having in mind its dependence on various parameters from Tables \ref{tab:crust} and
\ref{tab:subfaults}.

\subsection{Dynamic co-seismic displacements}\label{sec:dyndisp}

Here, we go even further in the reconstruction of the bottom motion. By making some assumptions on the time dependence of the displacement fields, we can have an insight into the dynamics of the sea bed motion.

The finite fault solution provides two additional parameters concerning the rupture dynamics for the July 17, 2006 event --- the rupture velocity $v_r = 1.1$ km/s and the rise time $t_r = 8$ s.  The epicenter is located at the point $\x_e =$ $(107.345^\circ, -9.295^\circ)$ \cite{Ji2006}. Given the origin $\x_e$, the rupture velocity $v_r$ and the $i^{\mathrm{th}}$ subfault location $\x_i$ (the full list is provided in Table \ref{tab:subfaults}), we define the {\em subfault activation times} $t_i$ needed for the rupture to reach the corresponding segment $i$ by the formula:
\begin{equation*}
  t_i = \frac{||\x_e - \x_i ||}{v_r}, \quad i=1,\ldots, N_x\times N_y.
\end{equation*}
For the sake of simplicity and due to the lack of information we assume implicitly that the rupture speed $v_r$ is constant along the fault; however this can be refined in future studies.

\begin{figure}
  \centering
  \includegraphics[width=0.69\textwidth]{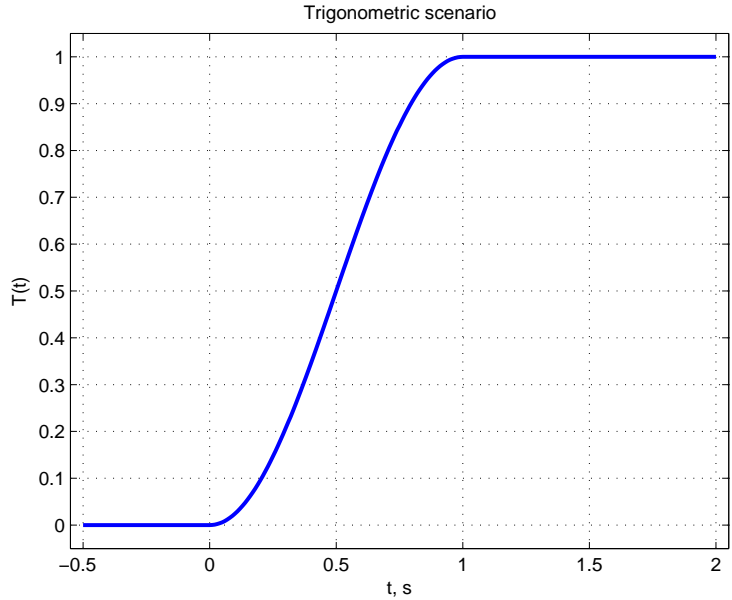}
  \caption{Trigonometric scenario with rise time $t_r = 1$ s.}
  \label{fig:trigscen}
\end{figure}

We will also follow the pioneering idea of J.~Hammack \cite{Hammack1972, Hammack} developed later in \cite{Todo, todo2, Dutykh2006, ddk, Kervella2007} where the maximum bottom deformation is achieved during some finite time (known as the rise time) according to a specific (in an \textit{ad hoc} manner) dynamic scenario. Various scenarios on the time dependence (instantaneous, linear, trigonometric, exponential, etc) can be found in \cite{Hammack, ddk, Dutykh2006}. In this study we will adopt the trigonometric scenario which can be described by the formula:
\begin{equation*}
  T(t) = \Hv(t-t_r) + \frac12\Hv(t)\Hv(t_r-t)\bigl(1 - \cos(\pi t/t_r)\bigr),
\end{equation*}
where $\Hv(t)$ is the Heaviside step function. For illustrative purposes this dynamic scenario is represented on Figure \ref{fig:trigscen}. Physically the function $T(t)$ represents the time history of the vertical bottom displacement in terms of its final amplitude. We assume that during the rise time temporal interval $[0, t_r]$ the vertical displacement goes from zero to its final stage according to the trigonometric scenario.

Finally, we put together all the ingredients in order to construct the dynamic sea bed motion:
\begin{equation}\label{eq:zeta}
  \zeta (\x, t) = \sum_{i=1}^{N_x\times N_y} \Hv(t-t_i) T(t-t_i) \O_i(\x).
\end{equation}
In the following sections we will present several approaches to couple this dynamic deformation with the hydrodynamic problem to predict waves induced on the ocean's free surface.

\section{Fluid layer solution}\label{sec:fluid}

Once the sea bed deformation is determined, a water wave problem must be solved in order to compute the free surface motion induced by the ocean bottom shaking. Traditionally this difficulty is circumvented by the simple translation of the static bottom deformation onto the free surface \cite{kajiura}, known as the passive generation approach \cite{ddk, Kervella2007}. In this section we present three approximate models to the water wave problem with moving bottom that we will use in combination with the finite-fault solution to study the tsunami generation problem.

\subsection{Linearized Euler equations -- CP model}\label{sec:CP}

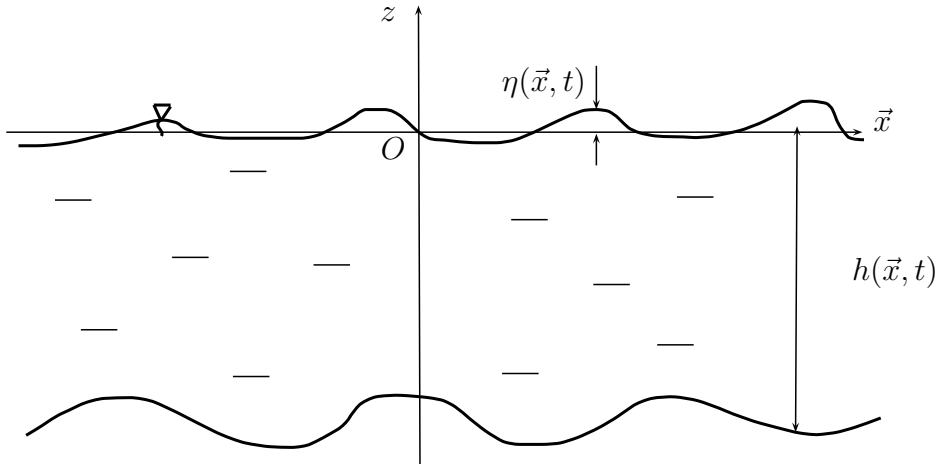
\begin{figure}
\begin{center}
\scalebox{1} 
{
\begin{pspicture}(0,-3.0692186)(12.901875,3.0992188)
\psline[linewidth=0.02cm,arrowsize=0.05291667cm 2.0,arrowlength=1.4,arrowinset=0.4]{->}(0.0,1.3407812)(11.26,1.3407812)
\psline[linewidth=0.02cm,arrowsize=0.05291667cm 2.0,arrowlength=1.4,arrowinset=0.4]{<-}(5.42,3.0207813)(5.44,-3.0592186)
\pscustom[linewidth=0.04]
{
\newpath
\moveto(0.16,1.1607813)
\lineto(0.246875,1.1607813)
\curveto(0.2903125,1.1607813)(0.3675348,1.1607813)(0.4013194,1.1607813)
\curveto(0.43510407,1.1607813)(0.5268056,1.1657813)(0.5847223,1.1707813)
\curveto(0.64263886,1.1757812)(0.75847226,1.1907812)(0.81638885,1.2007812)
\curveto(0.87430555,1.2107812)(0.9901388,1.2357812)(1.0480555,1.2507813)
\curveto(1.1059723,1.2657813)(1.2700695,1.3057812)(1.37625,1.3307812)
\curveto(1.4824306,1.3557812)(1.661007,1.4057813)(1.7334027,1.4307812)
\curveto(1.8057986,1.4557812)(1.9119793,1.4857812)(1.9457642,1.4907813)
\curveto(1.9795489,1.4957813)(2.037465,1.5007813)(2.0615973,1.5007813)
\curveto(2.0857294,1.5007813)(2.1291668,1.4957813)(2.1484723,1.4907813)
\curveto(2.1677778,1.4857812)(2.2063892,1.4757812)(2.2256947,1.4707812)
\curveto(2.245,1.4657812)(2.2787848,1.4457812)(2.293264,1.4307812)
\curveto(2.3077428,1.4157813)(2.3463545,1.3907813)(2.3704863,1.3807813)
\curveto(2.394618,1.3707813)(2.4525347,1.3457812)(2.4863195,1.3307812)
\curveto(2.5201044,1.3157812)(2.5925,1.2957813)(2.6311111,1.2907813)
\curveto(2.6697223,1.2857813)(2.742118,1.2757813)(2.7759027,1.2707813)
\curveto(2.8096876,1.2657813)(2.8869097,1.2607813)(2.9303472,1.2607813)
\curveto(2.9737847,1.2607813)(3.0558333,1.2607813)(3.0944445,1.2607813)
\curveto(3.1330554,1.2607813)(3.2102778,1.2607813)(3.248889,1.2607813)
\curveto(3.2875,1.2607813)(3.3695486,1.2607813)(3.412986,1.2607813)
\curveto(3.4564238,1.2607813)(3.5577776,1.2607813)(3.6156945,1.2607813)
\curveto(3.6736112,1.2607813)(3.8039236,1.2607813)(3.8763196,1.2607813)
\curveto(3.9487152,1.2607813)(4.083854,1.2907813)(4.1465974,1.3207812)
\curveto(4.20934,1.3507812)(4.339652,1.4157813)(4.407222,1.4507812)
\curveto(4.474791,1.4857812)(4.5761456,1.5507812)(4.6099305,1.5807812)
\curveto(4.6437154,1.6107812)(4.701632,1.6407813)(4.7257643,1.6407813)
\curveto(4.749896,1.6407813)(4.812639,1.6407813)(4.85125,1.6407813)
\curveto(4.8898606,1.6407813)(4.952604,1.6407813)(4.976736,1.6407813)
\curveto(5.000868,1.6407813)(5.049132,1.6307813)(5.073264,1.6207813)
\curveto(5.0973964,1.6107812)(5.150486,1.5807812)(5.179445,1.5607812)
\curveto(5.2084026,1.5407813)(5.275972,1.4807812)(5.3145833,1.4407812)
\curveto(5.3531947,1.4007813)(5.420764,1.3407812)(5.4497223,1.3207812)
\curveto(5.4786806,1.3007812)(5.531771,1.2707813)(5.5559034,1.2607813)
\curveto(5.580035,1.2507813)(5.637952,1.2407813)(5.671736,1.2407813)
\curveto(5.7055206,1.2407813)(5.816528,1.2307812)(5.89375,1.2207812)
\curveto(5.970972,1.2107812)(6.115764,1.2007812)(6.183334,1.2007812)
\curveto(6.250903,1.2007812)(6.381216,1.2007812)(6.4439588,1.2007812)
\curveto(6.5067015,1.2007812)(6.6321874,1.2207812)(6.6949306,1.2407813)
\curveto(6.7576733,1.2607813)(6.912118,1.3257812)(7.0038195,1.3707813)
\curveto(7.0955215,1.4157813)(7.2740974,1.5007813)(7.3609724,1.5407813)
\curveto(7.4478474,1.5807812)(7.568507,1.6257813)(7.602291,1.6307813)
\curveto(7.636076,1.6357813)(7.7132983,1.6407813)(7.756736,1.6407813)
\curveto(7.8001733,1.6407813)(7.8773956,1.6307813)(7.9111805,1.6207813)
\curveto(7.9449654,1.6107812)(8.017361,1.5607812)(8.055972,1.5207813)
\curveto(8.0945835,1.4807812)(8.157327,1.4207813)(8.181458,1.4007813)
\curveto(8.205591,1.3807813)(8.282812,1.3457812)(8.335903,1.3307812)
\curveto(8.388993,1.3157812)(8.490347,1.2957813)(8.53861,1.2907813)
\curveto(8.586875,1.2857813)(8.702709,1.2807813)(8.770278,1.2807813)
\curveto(8.837848,1.2807813)(8.963334,1.2757813)(9.02125,1.2707813)
\curveto(9.079166,1.2657813)(9.223959,1.2757813)(9.310834,1.2907813)
\curveto(9.397709,1.3057812)(9.561807,1.3407812)(9.639029,1.3607812)
\curveto(9.71625,1.3807813)(9.885174,1.4457812)(9.976875,1.4907813)
\curveto(10.068577,1.5357813)(10.2230215,1.6157813)(10.285764,1.6507813)
\curveto(10.348507,1.6857812)(10.430555,1.7307812)(10.449861,1.7407813)
\curveto(10.469166,1.7507813)(10.551214,1.7557813)(10.613957,1.7507813)
\curveto(10.676702,1.7457813)(10.7635765,1.7307812)(10.787707,1.7207812)
\curveto(10.81184,1.7107812)(10.850451,1.6707813)(10.86493,1.6407813)
\curveto(10.879409,1.6107812)(10.9035425,1.5507812)(10.913195,1.5207813)
\curveto(10.922847,1.4907813)(10.956632,1.4257812)(10.980764,1.3907813)
\curveto(11.004896,1.3557812)(11.048334,1.3007812)(11.067639,1.2807813)
\curveto(11.086945,1.2607813)(11.149688,1.2407813)(11.28,1.2407813)
}
\usefont{T1}{ptm}{m}{n}
\rput(11.521407,1.5107813){$\vec{x}$}
\usefont{T1}{ptm}{m}{n}
\rput(5.021406,2.9107811){$z$}
\pscustom[linewidth=0.04]
{
\newpath
\moveto(0.26,-2.6792188)
\lineto(0.36,-2.6192188)
\curveto(0.41,-2.5892189)(0.51,-2.5242188)(0.56,-2.4892187)
\curveto(0.61,-2.4542189)(0.695,-2.4042187)(0.73,-2.3892188)
\curveto(0.765,-2.3742187)(0.88,-2.3192186)(0.96,-2.2792187)
\curveto(1.04,-2.2392187)(1.205,-2.1942186)(1.29,-2.1892188)
\curveto(1.375,-2.1842186)(1.545,-2.1792188)(1.63,-2.1792188)
\curveto(1.715,-2.1792188)(1.875,-2.2092187)(1.95,-2.2392187)
\curveto(2.025,-2.2692187)(2.205,-2.3542187)(2.31,-2.4092188)
\curveto(2.415,-2.4642189)(2.63,-2.5742188)(2.74,-2.6292188)
\curveto(2.85,-2.6842186)(3.05,-2.7592187)(3.14,-2.7792187)
\curveto(3.23,-2.7992187)(3.39,-2.8242188)(3.46,-2.8292189)
\curveto(3.53,-2.8342187)(3.675,-2.8392189)(3.75,-2.8392189)
\curveto(3.825,-2.8392189)(3.945,-2.8192186)(3.99,-2.7992187)
\curveto(4.035,-2.7792187)(4.125,-2.7342188)(4.17,-2.7092187)
\curveto(4.215,-2.6842186)(4.31,-2.6292188)(4.36,-2.5992188)
\curveto(4.41,-2.5692186)(4.485,-2.4942188)(4.51,-2.4492188)
\curveto(4.535,-2.4042187)(4.6,-2.3192186)(4.64,-2.2792187)
\curveto(4.68,-2.2392187)(4.79,-2.1842186)(4.86,-2.1692188)
\curveto(4.93,-2.1542187)(5.1,-2.1442187)(5.2,-2.1492188)
\curveto(5.3,-2.1542187)(5.475,-2.1692188)(5.55,-2.1792188)
\curveto(5.625,-2.1892188)(5.725,-2.2092187)(5.75,-2.2192187)
\curveto(5.775,-2.2292187)(5.835,-2.2592187)(5.87,-2.2792187)
\curveto(5.905,-2.2992187)(5.995,-2.3692188)(6.05,-2.4192188)
\curveto(6.105,-2.4692187)(6.23,-2.5692186)(6.3,-2.6192188)
\curveto(6.37,-2.6692188)(6.485,-2.7392187)(6.53,-2.7592187)
\curveto(6.575,-2.7792187)(6.685,-2.7992187)(6.75,-2.7992187)
\curveto(6.815,-2.7992187)(6.96,-2.7992187)(7.04,-2.7992187)
\curveto(7.12,-2.7992187)(7.245,-2.7892187)(7.29,-2.7792187)
\curveto(7.335,-2.7692187)(7.445,-2.7342188)(7.51,-2.7092187)
\curveto(7.575,-2.6842186)(7.715,-2.6192188)(7.79,-2.5792189)
\curveto(7.865,-2.5392187)(7.99,-2.4642189)(8.04,-2.4292188)
\curveto(8.09,-2.3942187)(8.19,-2.3242188)(8.24,-2.2892187)
\curveto(8.29,-2.2542188)(8.415,-2.2042189)(8.49,-2.1892188)
\curveto(8.565,-2.1742187)(8.715,-2.1642187)(8.79,-2.1692188)
\curveto(8.865,-2.1742187)(9.015,-2.1992188)(9.09,-2.2192187)
\curveto(9.165,-2.2392187)(9.355,-2.3092186)(9.47,-2.3592188)
\curveto(9.585,-2.4092188)(9.81,-2.4892187)(9.92,-2.5192187)
\curveto(10.03,-2.5492187)(10.24,-2.6042187)(10.34,-2.6292188)
\curveto(10.44,-2.6542187)(10.61,-2.6742187)(10.68,-2.6692188)
\curveto(10.75,-2.6642187)(10.87,-2.6492188)(10.92,-2.6392188)
\curveto(10.97,-2.6292188)(11.095,-2.5942187)(11.17,-2.5692186)
\curveto(11.245,-2.5442188)(11.41,-2.4842188)(11.5,-2.4492188)
}
\psline[linewidth=0.02cm,arrowsize=0.05291667cm 2.0,arrowlength=1.4,arrowinset=0.4]{<->}(10.4,1.4207813)(10.38,-2.6392188)
\usefont{T1}{ptm}{m}{n}
\rput(11.691406,-0.50921875){$h (\vec{x}, t)$}
\usefont{T1}{ptm}{m}{n}
\rput(5.0914063,1.1307813){$O$}
\psline[linewidth=0.02cm,arrowsize=0.05291667cm 2.0,arrowlength=1.4,arrowinset=0.4]{<-}(7.76,1.3207812)(7.76,0.9007813)
\psline[linewidth=0.02cm,arrowsize=0.05291667cm 2.0,arrowlength=1.4,arrowinset=0.4]{<-}(7.76,1.6407813)(7.76,2.2207813)
\usefont{T1}{ptm}{m}{n}
\rput(7.0714064,1.9907813){$\eta(\vec{x}, t)$}
\psline[linewidth=0.02cm](0.64,0.44078124)(1.12,0.44078124)
\psline[linewidth=0.02cm](2.94,0.82078123)(3.42,0.82078123)
\psline[linewidth=0.02cm](2.98,-1.8992188)(3.46,-1.8992188)
\psline[linewidth=0.02cm](4.04,-0.41921875)(4.52,-0.41921875)
\psline[linewidth=0.02cm](0.98,-1.2792188)(1.46,-1.2792188)
\psline[linewidth=0.02cm](8.82,0.48078126)(9.3,0.48078126)
\psline[linewidth=0.02cm](8.56,-1.4792187)(9.04,-1.4792187)
\psline[linewidth=0.02cm](6.64,0.18078125)(7.12,0.18078125)
\psline[linewidth=0.02cm](6.52,-1.8592187)(7.0,-1.8592187)
\psline[linewidth=0.02cm](2.18,-0.31921875)(2.66,-0.31921875)
\psline[linewidth=0.02cm](7.72,-0.67921877)(8.2,-0.67921877)
\psline[linewidth=0.04cm](1.94,1.7007812)(2.04,1.4807812)
\psline[linewidth=0.04cm](2.04,1.5007813)(2.16,1.6807812)
\psline[linewidth=0.04cm](1.94,1.7007812)(2.16,1.7007812)
\pscustom[linewidth=0.04]
{
\newpath
\moveto(2.02,1.5007813)
\lineto(2.0,1.4707812)
\curveto(1.99,1.4557812)(1.99,1.4257812)(2.0,1.4107813)
\curveto(2.01,1.3957813)(2.03,1.3657813)(2.04,1.3507812)
\curveto(2.05,1.3357812)(2.05,1.3057812)(2.04,1.2907813)
}
\end{pspicture} 
}
\caption{Sketch of the physical domain.}
\label{fig:sketch}
\end{center}
\end{figure}

Consider an ideal incompressible fluid of constant density $\rho$. The horizontal projection of the fluid domain $\Omega$ is a subset of $\R^2$. The horizontal independent variables are denoted by $\x = (x,y)$ and the vertical one by $z$. The origin of the cartesian coordinate system is chosen such that the surface $z=0$ corresponds to the still water level. The fluid is bounded below by the bottom $z = -h(\x,t)$ and above by the free surface $z = \eta (\x,t)$. Usually we assume that the total depth $H(\x, t) := h(\x,t) + \eta (\x,t)$ remains positive $H (\x,t) \geq h_0 > 0$ at all times $t \in [0, T]$. The sketch of the physical domain is shown in Figure \ref{fig:sketch}.

\begin{remark}
Classically in water wave modeling, we make the assumption that the free surface is a graph $z = \eta (\x,t)$ of a single-valued function. It means in practice that we exclude some interesting phenomena, (e.g. wave breaking phenomena) which are out of the scope of this modeling paradigm.
\end{remark}

The linearized water wave problem consists of the following set of equations \cite{Hammack1972, Hammack, Dutykh2006}:
\begin{eqnarray}\label{eq:lin1}
  \Delta\phi = \grad^2\phi + \partial^2_{zz}\phi &=& 0, 
  \qquad (\x, z) \in \Omega\times [-h, 0], \\
  \dt\eta - \dz\phi &=& 0, \qquad z = 0, \\
  \dt\phi + g\eta &=& 0, \qquad z = 0, \\
  \dt h + \dz\phi &=& 0, \qquad z = -h(\x,t).\label{eq:lin2}
\end{eqnarray}
This set of equations together with an initial condition is also often referred to in the literature as the Cauchy-Poisson (CP) problem after the pioneering work of Cauchy \cite{Cauchy1827}.

In view of the specific requirements of the analytical techniques used in the applications, we will assume first that the domain $\Omega = \R^2$, i.e. it is unbounded in the horizontal extent, and the bottom has a special form:
\begin{equation*}
  h(\x, t) = h_0 - \zeta (\x,t),
\end{equation*}
where $h_0$ is some uniform depth and $\zeta(\x,t)$ is the sea bed displacement due to an underwater earthquake. In Section \ref{sec:dyndisp} one possible construction of the bottom displacement was proposed. Using integral transform methods (cf.  \cite{Hammack, Todo, Dutykh2006, Kervella2007}), one can derive the following expression for the free surface elevation $\eta(\x,t)$:
\begin{multline*}
  \eta(\x,t) = \frac{\gamma^2}{2}\F^{-1}\Bigl\{
  \sum_{i=1}^{n = N_x \times N_y}\frac{\Hv(t-t_i)\hat\O_i(\k)}{(\gamma^2 - \omega^2)\cosh(|\k|h_0)}
  \bigl(
  \cos(\omega(t-t_i)) - \cos(\gamma(t-t_i)) + \\ \Hv(t-t_i-t_r)[\cos(\omega(t-t_i-t_r)) + \cos(\gamma(t-t_i))]\bigr)
  \Bigr\},
\end{multline*}
where $t_r$ is the rise time defined in Section \ref{sec:dyndisp}, $\gamma = {\pi}/{t_r}$, 
$$ \omega^2 = g |\k| \tanh (|\k|h_0) $$ and $\F^{-1}$ is the inverse Fourier transform (see equation \eqref{eq:fourier} below). A similar expression can also be derived for the velocity potential $\phi (\x,z,t)$, however we do not directly need it in our study.

This analytical solution will be used below in numerical simulations. It has the advantage of being simple and, thus, computationally inexpensive. However, the flat bottom assumption ($h(\x) = h_0 = const$) prevents us from using this solution beyond some small evolution times. The validity of this approximation has already been addressed in the literature \cite{Kervella2007, Saito2009} and will be discussed at some point below.

\subsection{The weakly nonlinear (WN) model}\label{sec:WN}

A tsunami wave during its generation is usually well described by the Cauchy-Poisson problem \cite{Dutykh2006, Kervella2007, Saito2009}. The main reason for this simplification is the fact that a wave of a half meter amplitude represents only a tiny perturbation of over a 4000 m water column. However, the real world bathymetry is generally complex and may contain simultaneously various scales. For example, the subduction zone bathymetry represented on Figure \ref{fig:Java_fault} ranges from 7000 to 20 m and thus, nonlinear effects may be locally important. In order to take into account all realistic bathymetric features and study in detail the initial stages of tsunami propagation we describe below a new numerical model.

We consider the physical setting and notation of Section \ref{sec:CP}. The governing equations of the classical water wave problem are the following \cite{Lamb1932, Stoker1958, Mei1994, Whitham1999}:
\begin{eqnarray}
  \Delta\phi = \grad^2\phi + \partial^2_{zz}\phi &=& 0, 
  \qquad (\x, z) \in \Omega\times [-h, \eta], \label{eq:laplace} \\
  \dt\eta + \grad\phi\cdot\grad\eta - \dz\phi &=& 0, 
  \qquad z = \eta(\x, t), \label{eq:kinematic} \\
  \dt\phi + \half|\grad\phi|^2 + \half(\dz\phi)^2 + g\eta &=& 0, 
  \qquad z = \eta(\x,t), \label{eq:bernoulli} \\
  \dt h + \grad\phi\cdot\grad h + \dz\phi &=& 0, 
  \qquad z = -h(\x,t), \label{eq:bottomkin}
\end{eqnarray}
with $\phi$ the velocity potential, $g$ the acceleration due to gravity force and $\grad = (\partial_x, \partial_y)$ denotes the gradient operator in horizontal Cartesian coordinates.

The assumptions of fluid incompressibility and flow irrotationality lead to the Laplace equation (\ref{eq:laplace}) for the velocity potential $\phi(\x, z, t)$. The main difficulty of the water wave problem lies on the boundary conditions. Equations (\ref{eq:kinematic}) and (\ref{eq:bottomkin}) express the free-surface kinematic condition and bottom impermeability respectively, while the dynamic condition (\ref{eq:bernoulli}) expresses the free surface isobarity.

The bathymetry $h(\x,t)$ is decomposed into the static part $h_0(\x)$ (given e.g.  by the ETOPO1 database, cf. Figure \ref{fig:Java_fault}) and the dynamic sea bed displacement $\zeta(\x,t)$ constructed above in \eqref{eq:zeta}:
\begin{equation}\label{eq:bottom}
  h(\x,t) = h_0 (\x) - \zeta (\x,t).
\end{equation}


\begin{remark}
Recently, some weak dissipative effects have also beed included into the classical water wave problem (\ref{eq:laplace}) -- (\ref{eq:bottomkin}). For more details on the visco-potential formulation we refer to \cite{Dias2007, DutykhDias2007, Dutykh2007a, Dutykh2008a, Dutykh2008b}.
\end{remark}

In the sequel we will need the unit exterior normals to the fluid domain. It is straightforward to obtain the following expressions for the normals at the free surface and bottom respectively:
\begin{equation*}
  \n_f = \frac{1}{\sqrt{1 + |\grad\eta|^2}} [-\grad\eta, 1]^t,
  \qquad
  \n_b = \frac{1}{\sqrt{1 + |\grad h|^2}} [-\grad h, -1]^t.
\end{equation*}

In 1968 Zakharov proposed a different formulation of the water wave problem based on the trace of the velocity potential at the free surface \cite{Zakharov1968}:
\begin{equation}\label{eq:trace}
  \phis(\x, t) := \phi(\x, \eta(\x,t), t).
\end{equation}
This variable plays the role of generalized momentum in the Hamiltonian description of water waves \cite{Zakharov1968, Dias2006a}. The second canonical variable is the free surface elevation $\eta$.

Another important ingredient is the normal velocity at the free surface $v_n$ which is defined as:
\begin{equation}\label{eq:normalv}
  v_n (\x,t) := \sqrt{1 + |\grad\eta|^2}\left.\pd{\phi}{\n_f}\right|_{z=\eta} = 
  \left.(\dz\phi - \grad\phi\cdot\grad\eta)\right|_{z=\eta}.
\end{equation}
The boundary conditions (\ref{eq:kinematic}) and (\ref{eq:bernoulli}) on the free surface can be rewritten in terms of $\phis$, $v_n$ and $\eta$ \cite{Craig1992, Craig1993, Fructus2005}:
\begin{equation}\label{eq:dynamics}
 \begin{array}{rl}
  \dt\eta - \D_\eta(\phis) &= 0, \\
  \dt\phis + \half|\grad\phis|^2 + g\eta
  - \frac{1}{2(1+|\grad\eta|^2)}\bigl[\D_\eta(\phis) + 
   \grad\phis\cdot\grad\eta \bigr]^2 &= 0.
 \end{array}
\end{equation}
Here we introduced the Dirichlet-to-Neumann operator (D2N) $\D_\eta : \phis \mapsto v_n$ \cite{Coifman1985, Craig1993} which maps the velocity potential at the free surface $\phis$ to the normal velocity $v_n$.
The name of this operator comes from the fact that it denotes a correspondance between Dirichlet data $\phis$ and Neumann data $\sqrt{1+|\grad\eta|^2}\left. \displaystyle{\pd{\phi}{\n_f}} \right|_{z=\eta}$ on the free surface. We provide in Appendix \ref{app:ii} the complete derivation of Zakharov's formulation for the water wave problem.

\subsubsection{Numerical evaluation of the D2N operator}

We saw above that the water wave problem can be reduced to a system of two PDEs governing the evolution of the canonical variables $\eta$ and $\phis$. In order to solve this system of equations we must be able to compute efficiently the quantity $\D_\eta(\phis)$. In this section we present a simple method for the numerical computation of the D2N operator, which is appropriate for the application of the linearized Euler model in the solution of problems dealing with tsunami generation. This approach is based on the extensive use of Fourier transforms. On the discrete level this transformation can be efficiently implemented with the Fast Fourier Transform (FFT) algorithm \cite{Cooley1965, Frigo2005}.

The direct $\F$ and inverse $\F^{-1}$ Fourier transforms in 2D are defined as follows:
\begin{equation}\label{eq:fourier}
  \F[f] = \hat{f}(\k) = \int\limits_{\R^2} f(\x) e^{-i\k\cdot\x}\; d\x, \quad
  \F^{-1}[\hat{f}] = f(\x) = \frac{1}{(2\pi)^2}\int\limits_{\R^2} \hat{f}(\k)
  e^{i\k\cdot\x}\; d\k.
\end{equation}

The problem to be solved is
\begin{eqnarray}\label{eq:lapl}
	\grad^2\phi + \partial^2_{zz}\phi &=& 0, \quad (\x, z) \in \Omega\times [-h, \eta], \\
  \phi &=& \phis, \quad z=\eta, \label{eq:dirichlet} \\
  \sqrt{1 + |\grad h|^2}\displaystyle{\pd{\phi}{\n_b}} &=& \dt h, \quad z=-h.
  \label{eq:neumann}
\end{eqnarray}
Once the function $\phi$ is determined, we must compute its normal derivative on the free surface (\ref{eq:normalv}).  

Since a tsunami wave induces a special flow regime in which the horizontal extent is much more important than the variations in the vertical direction, we can apply the Fourier transform to the Laplace equation (\ref{eq:lapl}) as if it were posed in a strip-like domain:
\begin{equation*}
  \od{^2\phih}{z^2} - |\k|^2\phih = 0.
\end{equation*}
The general exact solution to this ODE can be easily computed:
\begin{equation}\label{eq:gensol}
  \phih(\k; z) = A(\k)\cosh(|\k|z) + B(\k)\sinh(|\k|z).
\end{equation}
The two unknown functions $A(\k)$ and $B(\k)$ must be determined from the boundary conditions (\ref{eq:dirichlet}), (\ref{eq:neumann}). For the sake of convenience we rewrite the Neumann boundary condition at the bottom (\ref{eq:neumann}) in this form:
\begin{equation}\label{eq:forcing}
  \left.\pd{\phi}{z}\right|_{z=-h} = -\dt h - \left.\grad\phi\right|_{z=-h}\cdot\grad h 
  \equiv f(\x, t).
\end{equation}
The right-hand side will be denoted by $f(\x,t)$, which implicitly depends on the solution $\phi$.

The application of the boundary conditions leads to the following system of linear equations:
\begin{eqnarray*}
  \cosh(|\k|\eta)A(\k) + \sinh(|\k|\eta)B(\k) &=& \hat{\phis} \\
  -|\k|\sinh(|\k|h)A(\k) + |\k|\cosh(|\k|h)B(\k) &=& \hat{f},
\end{eqnarray*}
which can be easily solved:
\begin{equation*}
  A(\k) = \frac{\hat{\phis}\cosh(|\k|h) -
            \hat{f}\displaystyle{\frac{\sinh(|\k|\eta)}{|\k|}}}{\cosh(|\k|H)}, \quad
  B(\k) = \frac{\hat{\phis}\sinh(|\k|h) + 
  			    \hat{f}\displaystyle{\frac{\cosh(|\k|\eta)}{|\k|}}}{\cosh(|\k|H)}.
\end{equation*}
Here, $H = h + \eta$ is the total water depth. The knowledge of these functions provides the velocity potential in the whole domain thanks to the general solution (\ref{eq:gensol}).

Finally, we compute the normal velocity $v_n$ on the free surface (\ref{eq:normalv}).
If we compute this quantity in Fourier space, the answer will be given immediately by the inverse transform $\F^{-1}$. The first term  of $v_n$ is readily given by the formula
\begin{equation*}
  \left.\dz\phih\right|_{z=\eta} = \hat{\phis}|\k|\tanh(|\k|H) + \hat{f}\sech(|\k|H).
\end{equation*}
To compute the second term we use the following approximate expression:
\begin{equation}\label{eq:nlinproduct}
  \widehat{\left.\grad\phi\right|_{z=\eta}\cdot\grad\eta} =  \F\Bigl[\F^{-1}\bigl[i\k\hat{\phis}\bigr]\cdot\F^{-1}\bigl[i\k\hat{\eta}\bigr]\Bigr].
\end{equation}

\begin{remark}
Equation \eqref{eq:forcing} indicates that the function $f(\x,t_n)$ depends implicitly on the unknown solution $\phi(\x, z, t_n)$. In order to compute this apparent contradiction, we apply a fixed-point iteration initialized with the value of $f(\x,t_{n-1})$ from the previous time step:
\begin{equation*}
  \hat{f}^{k+1} = -\widehat{\dt h} - \F\Bigl[\left.\grad\phi\right|_{z=-h}(\hat{f}^k)
  \cdot\grad h\Bigr], \quad \hat{f}^0 = \hat{f}(\k; t_{n-1}).
\end{equation*}
The last product is computed in the physical space:
\begin{equation*}
  \F\Bigl\{\left.\grad\phi\right|_{z=-h}(f^k)\cdot\grad h\Bigr\} = \F\Bigl[
    \F^{-1}\bigl[\left.\widehat{\grad\phi}\right|_{z=-h}(\hat{f}^k)\bigr]\cdot
    \grad h\Bigr].
\end{equation*}
Simple computations yield
\begin{equation*}
  \left.\widehat{\grad\phi}\right|_{z=-h}(\hat{f}^k) = i\k\Bigl[
    \hat{\phis}\sech(|\k|H) - \hat{f}^k\frac{\tanh(|\k|H)}{|\k|}
  \Bigr].
\end{equation*}
Our numerical experiments show that this iterative procedure is convergent and the tolerance $\varepsilon := ||\hat{f}^{k+1} - \hat{f}^k||_\infty \leq 10^{-5}$ is reached after four iterations in average.
\end{remark}

The resulting model is only weakly nonlinear since Laplace's equation \eqref{eq:lapl} is solved using the Fourier transform in a strip-like domain. Consequently, there is an implicit linearization in the solution procedure. However, the WN model contrary to the CP model not only takes into account some nonlinear effects but can also be efficiently applied to cases with realistic bathymetry. We note that this model is similar to the first order approximation model proposed in \cite{Guyenne2007} if in our method we further simplify all expressions by replacing the total water depth $H$ by the undisturbed depth $h$.

\subsection{Time integration}

Applying the above Fourier type spectral method to equations \eqref{eq:dynamics} governing the evolution of the canonical variables $\eta$ and $\phis$ leads to a system of ordinary differential equations, i.e.
\begin{equation}\label{ode1}
  \Phi_t = \A(t, \Phi), \qquad \Phi(t_0) = \Phi_0, \qquad \Phi = (\eta,\phis)^T.
\end{equation}
In order to integrate numerically this system of ODEs we apply an integrating factor method analogous to the one used in \cite{Fructus2005, Xu2009}. This method apparently decreases the stiffness of the system of ODEs and therefore allows for an efficient application of explicit time integration schemes. We start by extracting the linear part of equations (\ref{ode1}):
\begin{equation}\label{ode2}
  \Phi_t + \L\cdot\Phi = \N(\Phi),
\end{equation}
where $  \L = \begin{pmatrix}
         0 & -\frac{\omega^2}{g} \\
         g & 0 \\
       \end{pmatrix}$ and $\omega = \sqrt{g|\k|\tanh(|\k|h_0)}$ is the wave frequency corresponding to the wave number $|\k|$. For a general bathymetry we choose the constant $h_0$ to be the mean water depth. (We note that we use the arithmetic average of values provided by the ETOPO1 database in the region under consideration.) The term $\N(\Phi)$ incorporates the remaining nonlinear terms:
\begin{equation*}
  \N(\Phi) = \begin{pmatrix}
    \F\bigl\{\D_\eta(\phis)\bigr\} - \frac{\omega^2}{g}\hat{\phis} \\
    \F\Bigl\{\frac{1}{2(1+|\grad\eta|^2)}\bigl[\D_\eta(\phis) + 
   \grad\phis\cdot\grad\eta \bigr]^2 - \half|\grad\phis|^2\Bigr\}
  \end{pmatrix}.
\end{equation*}
The linear terms can be integrated exactly by the following change of variables:
\begin{equation*}
  \Psi(t) := e^{\L(t-t_0)}\Phi(t), \qquad
  e^{\L(t-t_0)} = \begin{pmatrix}
    \cos(\omega(t-t_0)) & -\frac{\omega}{g}\sin(\omega(t-t_0)) \\
    \frac{g}{\omega}\sin(\omega(t-t_0))  & \cos(\omega(t-t_0))
  \end{pmatrix}.
\end{equation*}
Consequently, we solve in practice the following system of ODEs:
\begin{equation*}
  \Psi_t = e^{\L(t-t_0)}\N\bigl(e^{-\L(t-t_0)}\Psi\bigr) \equiv \B(t,\Psi), \qquad
  \Psi(t_0) = \Phi_0.
\end{equation*}
This simple modification allows us to take larger CFL numbers, thus improving the overall time stepping performance.

Finally, the system of ODEs is discretized by the standard fourth-order Runge-Kutta (RK4) scheme \cite{Hairer2009}:
\begin{equation}\label{eq:rk4}
  \begin{array}{rl}
  	\Psi_{n+1} &= \Psi_n + \frac16 \Delta t (k_1 + 2k_2 + 2k_3 + k_4), \\
  	k_1 &= \B (t_n, \Psi_n), \\
  	k_2 &= \B (t_n + \frac12 \Delta t, \Psi_n + \frac12 \Delta t \; k_1), \\
  	k_3 &= \B (t_n + \frac12 \Delta t, \Psi_n + \frac12 \Delta t \; k_2), \\
  	k_4 &= \B (t_n + \Delta t, \Psi_n + \Delta t \; k_3),
  \end{array}
\end{equation}
where the subscript refers to the discrete time instance  $\Psi_n := \Psi (t_n)$ and $\Delta t$ is the discrete time step: $t_{n+1} = t_n + \Delta t$.

In the computations described below, we use a Runge-Kutta (4,5) scheme with an adaptive time step control (cf. \cite{Dormand1980}). However it is not so fundamentally different from the classical RK4 scheme \eqref{eq:rk4} described above.

\subsection{The BBM-BBM type system}

When the long wave approximation is applied to the water wave problem \eqref{eq:laplace} -- \eqref{eq:bottomkin}, one obtains the well-known nonlinear shallow water (or Saint-Venant) equations \cite{SV1871, Stoker1958, Whitham1999} which have been extensively used for tsunami simulations \cite{Imamura1996, Titov1997, TS, DeKaKa, Dutykh2009a}. If we go further in the asymptotic expansions, some dispersive effects can be included and generally the resulting system is referred to as Boussinesq system \cite{Boussinesq1872, BCS, Madsen03, Dutykh2007, DMS1, DMS2}.

In this study we use the Boussinesq system of BBM-BBM type with variable bottom derived in \cite{Mitsotakis2007}. See also \cite{Peregrine1967, Chazel2007}. The system in dimensional variables can be written as:
\begin{equation}\label{eq:Bouss}
\begin{array}{rl}
 \eta_t + \div((h_0+\eta)\u) + \div\left\{A h_0^2[\nabla(\grad h_0\cdot\u) + 
 \grad h_0\div\u] - bh_0^2\grad\eta_t\right\} + & \\
 A\nabla\cdot(h_0^2\nabla\zeta_t) + \zeta_t & = 0, \\
 \u_t + g\grad\eta + \frac12 \grad|\u|^2 + B g h_0[\grad(\grad h\cdot\nabla\eta) + \grad h_0\Delta\eta] - d h_0^2\Delta\u_t
 - B h_0\grad\zeta_{tt} &= 0,
\end{array}
\end{equation}
where $A$, $B$, $b$ and $d$ are constants defined as:
\begin{equation*}
  A = \sqrt{\frac{2}{3}}-\frac{2}{3}, \quad
  B = 1-\sqrt{\frac{2}{3}}, \quad 
  b = d = \frac{1}{6}.
\end{equation*}
The variable $\u(\x,t)$ denotes the horizontal velocity of the fluid at $z=-h+\sqrt{2/3}(\eta+h)$, and the bathymetry variables $h(\x,t)$, $h_0(\x)$, $\zeta (\x,t)$ are defined in Section \ref{sec:displ}.

We integrate numerically the system \eqref{eq:Bouss} by using the standard Galerkin/finite element method with $\P1$ elements for the spatial discretization coupled with an explicit, second-order Runge-Kutta method for the temporal discretization (so-called improved Euler scheme) \cite{Hairer2009}. A proof that the semidiscrete system is not stiff and thus that the specific RK method is sufficient can be found in \cite{Dougalis2010}.

In order to obtain a well-posed problem, we impose homogeneous Dirichlet boundary conditions which absorb partially the wave while reflecting only small amplitude oscillatory waves. Moreover, the specific numerical method appears to converge with optimal rate in the $L^2$ and $L^\infty$ norms whether we consider structured or unstructured grids. This is contrary to the analogous initial boundary value problems with zero Dirichlet boundary conditions on $\u$ for the Peregrine system \cite{Peregrine1967} where the analogous numerical method converges with suboptimal orders on structured and unstructured grids. For more information on the properties and the implementation of the numerical method for a BBM-BBM type system we refer to \cite{DMS1, Mitsotakis2007}.

\section{Numerical results}\label{sec:numres}

In this section we compare the propagation of a solitary wave when it is used as an initial condition in both the CP and WN models. Moreover, we study the generation and the initial stages of the propagation of the tsunami wave of the July 17, 2006 event. We also present a comparison between the WN, CP and Boussinesq models.

\subsection{Solitary wave propagation}

Before performing the Java 2006 tsunami generation simulations, we study the propagation of a solitary wave solution to the full water wave problem using the WN and CP models. The initial condition is a solitary wave, computed by using the method presented by Tanaka \cite{Tanaka1986}.

Consider the  two-dimensional water wave problem in a channel of uniform depth $h_0 = const$. Since we look for travelling wave solutions, the flow field can be reduced to the steady state by choosing a frame of reference moving with the wave speed $c$. The introduction of dimensionless variables leads to a single scaling parameter, the Froude number $\Fr$ defined as $\Fr := \displaystyle{\frac{c}{\sqrt{gh_0}}}$. Hereafter, the governing equations are considered in dimensionless form.

The complex velocity potential is classically introduced as $w = \phi + i\psi$, where $\psi$ is the stream function. We choose $\phi = 0$ at the crest and $\psi = 0$ at the bottom. The fluid region is then mapped onto the strip $0<\psi<1$, $-\infty<\phi<\infty$ on the plane $w$ with $\psi = 1$ corresponding to the free surface. We introduce the quantity $\Omega = \log\displaystyle{\od{w}{z}} = \tau - i\theta$, where $\theta$ is the angle between the velocity vector and horizontal axis $Ox$. The real part $\tau$ is expressed in terms of the velocity magnitude $q$ as $\tau = \log q$. The boundary conditions to be satisfied are the dynamic condition on the free surface and the bottom impermeability which are expressed as
\begin{equation}\label{eq:bcond}
  \od{q^3}{\phi} = -\frac{3}{\Fr^2}\,\sin\theta, \quad\mbox{on } \psi = 1
  \qquad \mbox{ and } \qquad \theta = 0, \quad \mbox{on } \psi = 0.
\end{equation}

Consequently, the problem is now transformed into the determination of the complex function $\Omega$, analytic with respect to $w$ within the region of the unit strip $0 < \psi < 1$, decaying at infinity and satisfying the boundary conditions (\ref{eq:bcond}). By applying Cauchy's integral theorem, one can find the following integral equation on the free surface $\psi = 1$:
\begin{equation*}
  -\theta(\phi) 
  - \frac{2}{\pi} 
  \int\limits_{-\infty}^\infty\frac{\theta(\phi)}{(\phis-\phi)^2 + 4}\;d\phis 
  = -\frac{1}{\pi}\int\limits_{-\infty}^\infty
  \frac{(\phis - \phi)\tau(\phis)}{(\phis - \phi)^2 + 4}\; d\phis + 
  \frac{1}{\pi}\mbox{p.v.}\int\limits_{-\infty}^\infty
  \frac{\tau(\phis)}{\phis-\phi}\; d\phis,
\end{equation*}
where $\tau(\phi)$ and $\theta(\phi)$ denote the traces of the corresponding functions on the free surface $\psi = 1$.

\begin{figure}
  \centering
	\includegraphics[width=0.85\textwidth]{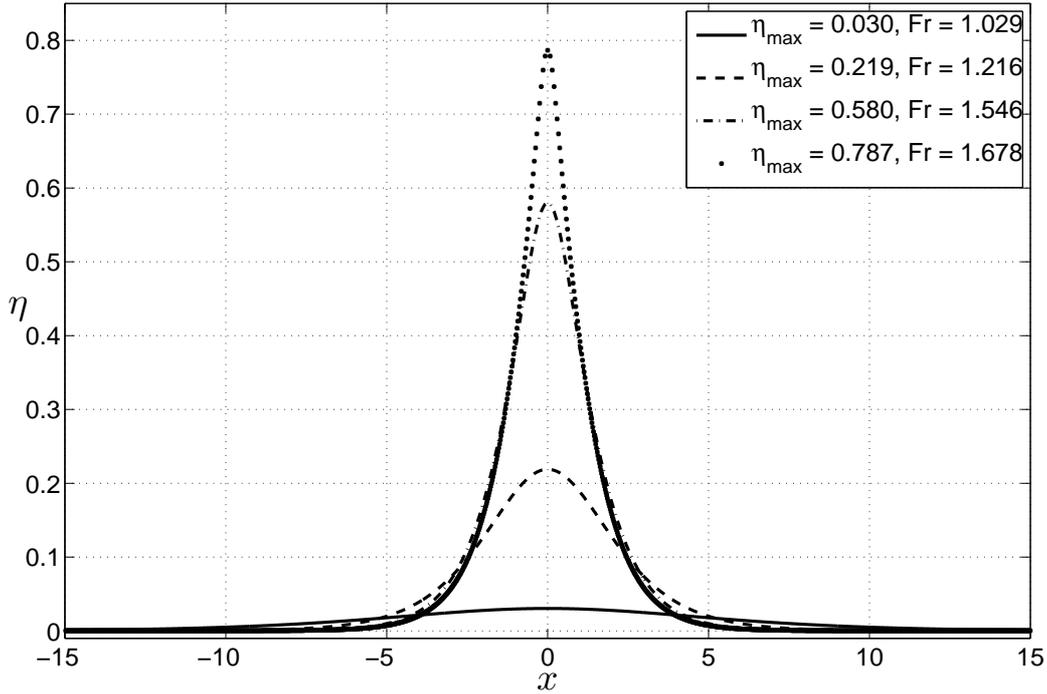}
  \caption{Solitary wave solutions of various amplitudes for the full water
	wave problem. Both $x$ and $\eta$ have been non-dimensionalized by the depth $h_0$.}
  \label{fig:tanaka}
\end{figure}

The integral equation is solved iteratively. The convergence is tested with respect to the Froude number. Several solitary wave solutions computed in this way are plotted on Figure \ref{fig:tanaka} for illustrative purposes.

In order to illustrate the advantages of the proposed WN model over the classical CP solution, we let a solitary wave with amplitude $A/h_0 = 0.1$ propagate up to $T = 80$ (in this section we use dimensionless quantities and time $T$ is non-dimensionalized by $\sqrt{g/h_0}$).

We recall that the classical CP solution of \eqref{eq:lin1}--\eqref{eq:lin2} corresponding to the initial free surface height $\left.\eta\right|_{t=0} = \eta_0(x)$ and the velocity potential distribution at the free surface $\left.\phis\right|_{t=0} = \phis_0(x)$, takes the following form:
\begin{equation*}
  \eta (\x,t) = \F^{-1}\Bigl\{\hat\eta_0(\k)\cos(\omega t) + 
  \frac{\omega}{g}\hat\phis_0(\k)\sin(\omega t)\Bigr\},
\end{equation*}
\begin{equation*}
  \phi (\x, z, t) = \F^{-1}\Bigl\{\bigl(
  \hat\phis_0(\k)\cos(\omega t) 
  - \frac{g}{\omega}\hat\eta_0(\k)\sin(\omega t)
  \bigr)
  \bigl(\cosh(|\k|z) + 
  \tanh(|\k|h)\sinh(|\k|z)\bigr)\Bigr\},
\end{equation*}
where $\hat\eta_0(\k) = \F\{\eta_0(\x)\}$ and $\hat\phis_0(\k) = \F\{\phis_0(\x)\}$ are the Fourier transforms of the initial conditions.

\begin{figure}%
\centering
\subfigure[$t=20$]{\includegraphics[scale=0.25]{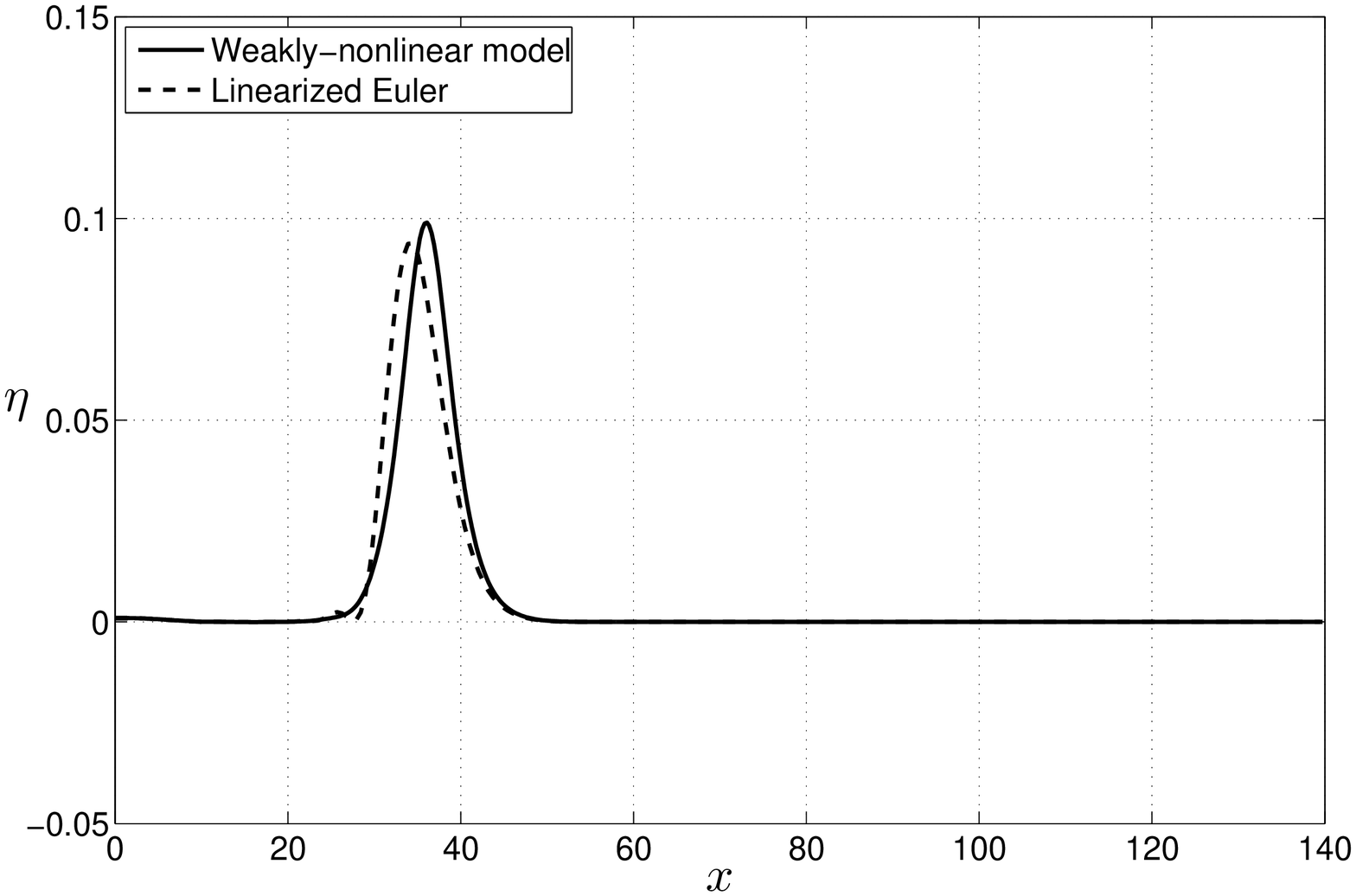}}
\subfigure[$t=40$]{\includegraphics[scale=0.25]{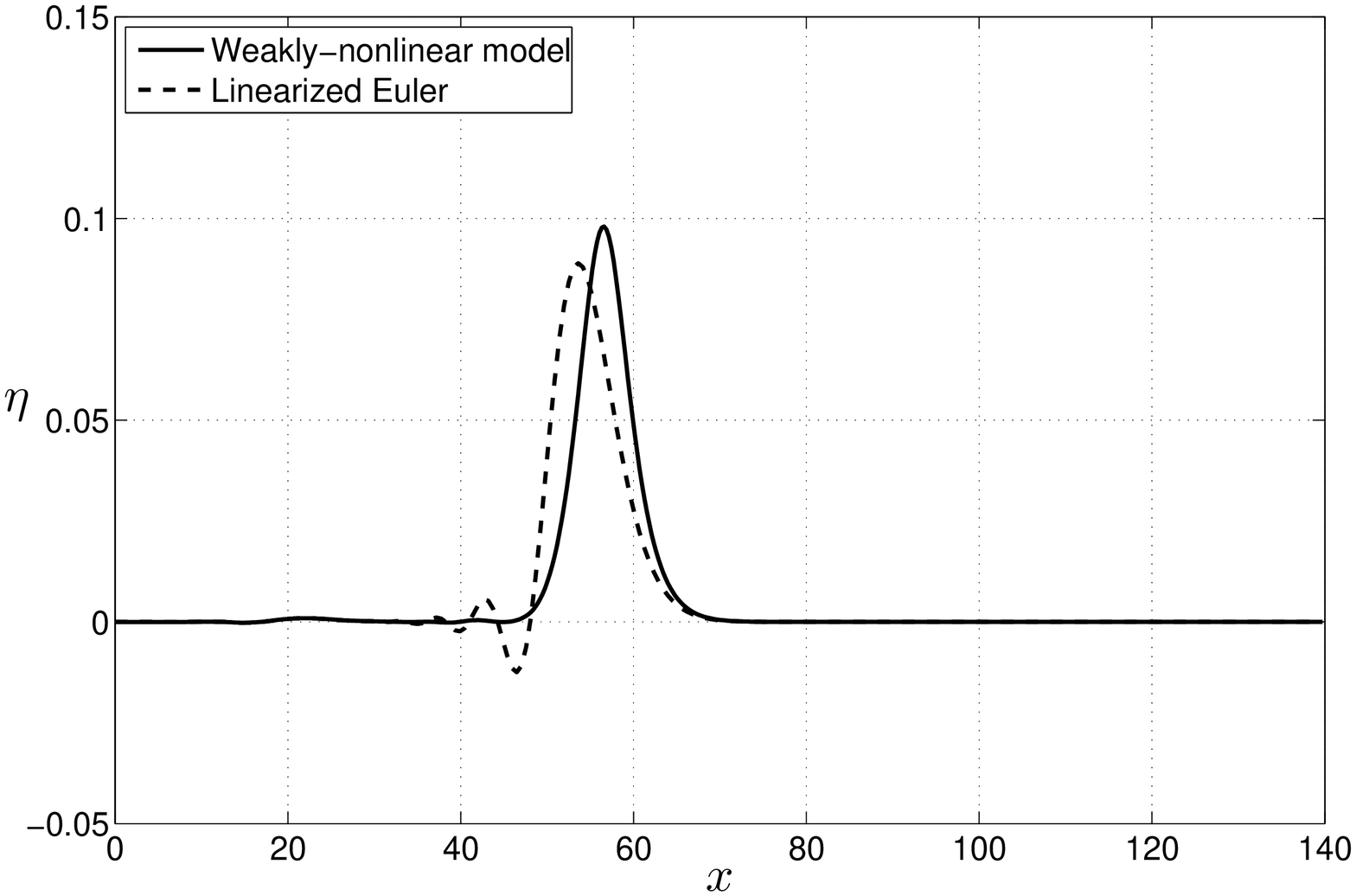}}
\subfigure[$t=60$]{\includegraphics[scale=0.25]{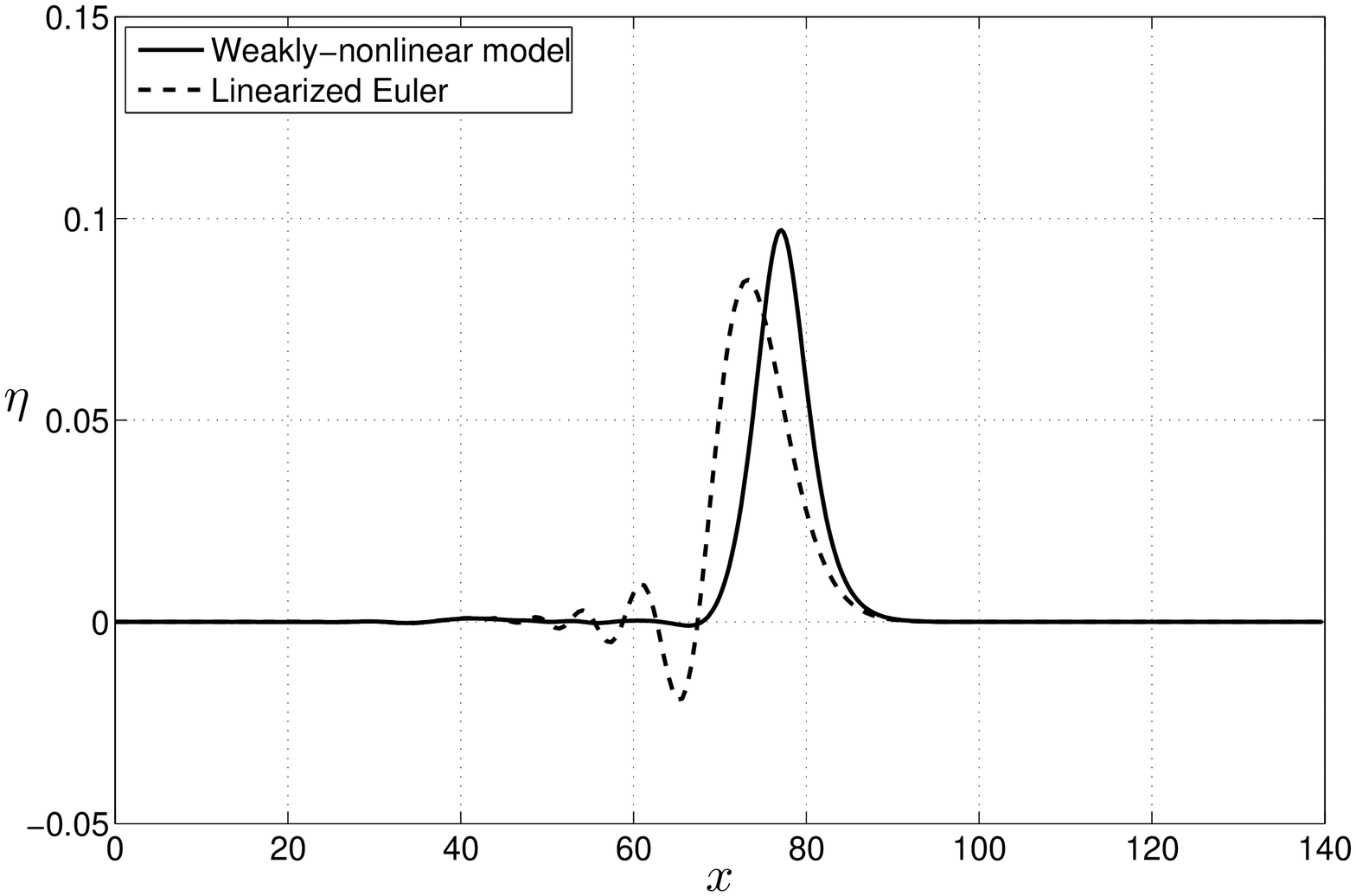}}
\subfigure[$t=70$]{\includegraphics[scale=0.25]{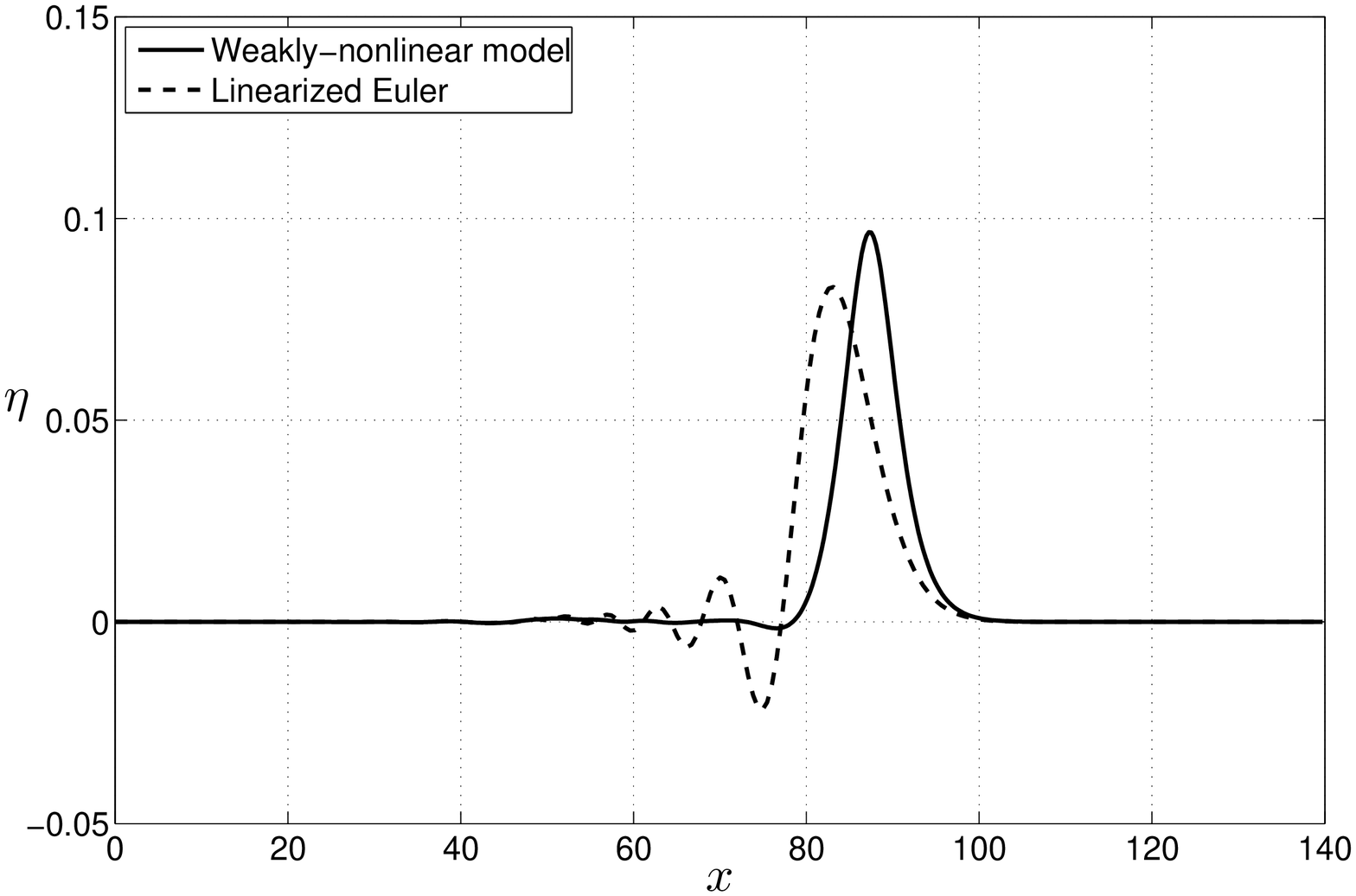}}
\subfigure[$t=80$]{\includegraphics[scale=0.25]{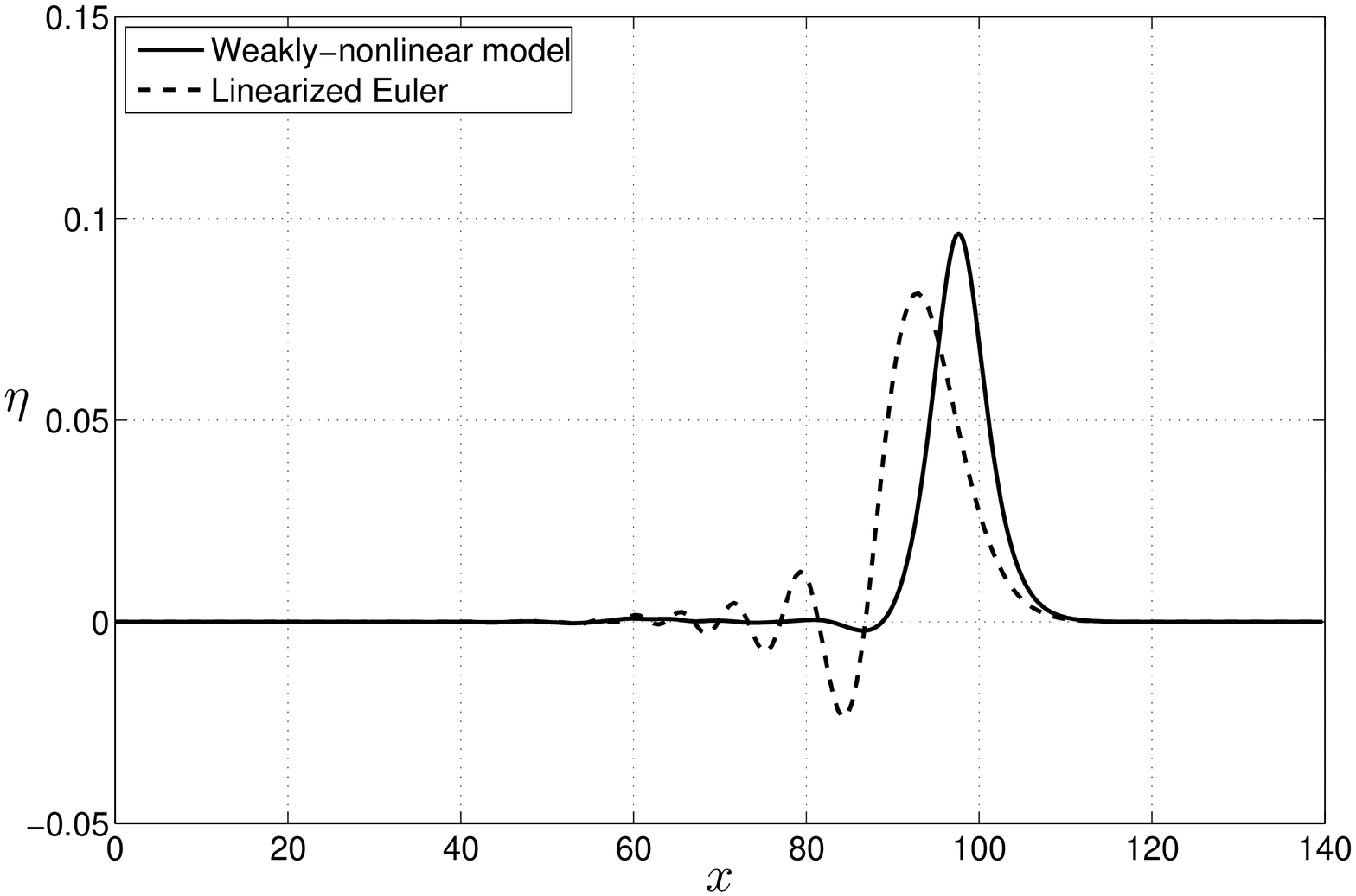}}
\subfigure[Amplitude error]{\includegraphics[scale=0.25]{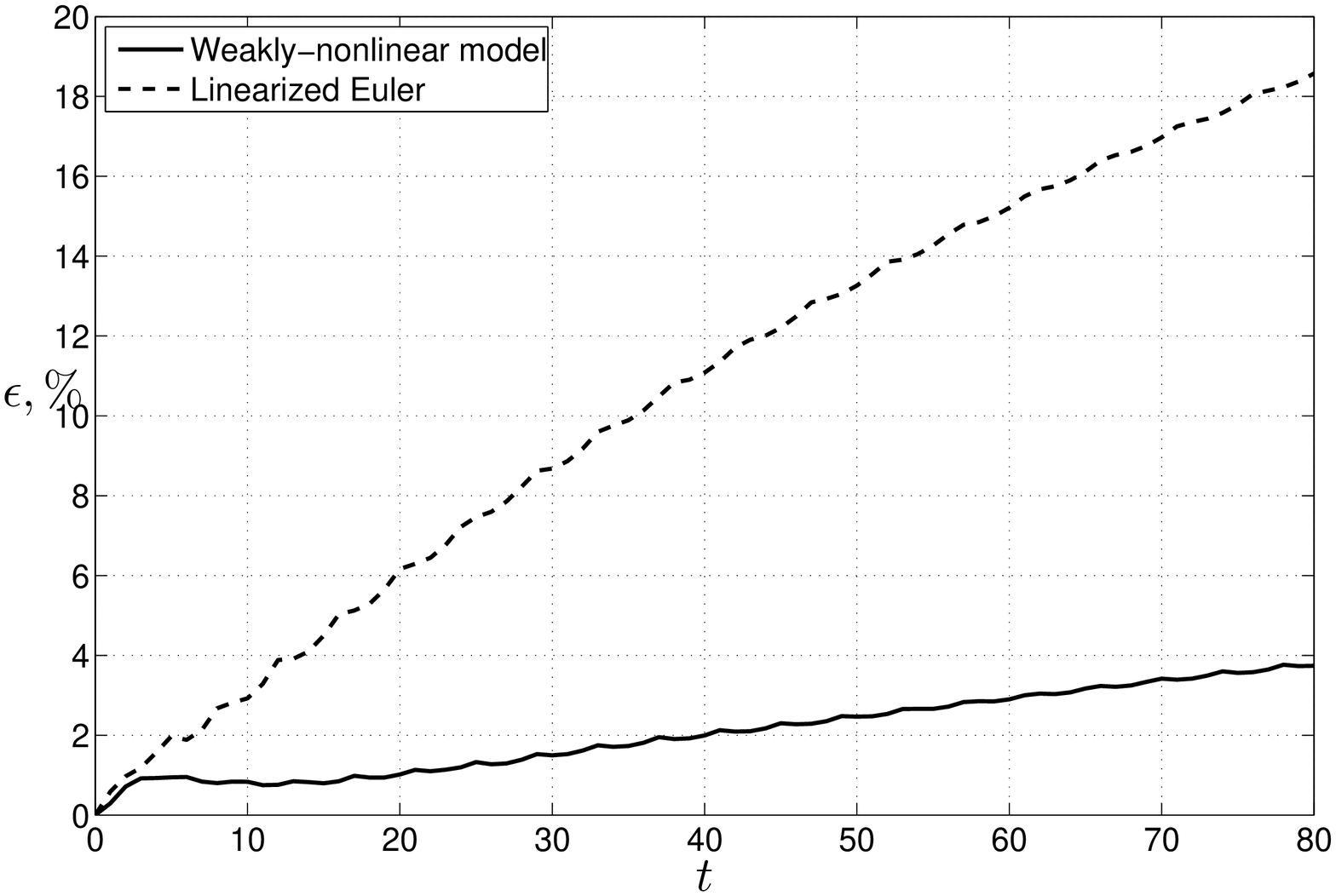}}
\caption{Propagation of a solitary wave with the weakly nonlinear method (solid line) and Cauchy-Poisson solution (dashed line). The solitary wave amplitude is $A/h_0 = 0.1$. Space has been scaled by $h_0$ and time by $\sqrt{g/h_0}$.}
\label{fig:tansev}%
\end{figure}

The solution profiles of both models are presented in Figures \ref{fig:tansev} (a)--(e). One observes that the WN model preserves quite well the shape of the solitary wave while shedding a small dispersive tail behind. The CP solution gradually transforms the initial wave into a dispersive tail according to the linear nature of equations \eqref{eq:lin1}--\eqref{eq:lin2}. In Figure \ref{fig:tansev} (f) we present the normalized amplitude error defined as:
\begin{equation*}
  \epsilon (t) := \frac{|\max\limits_x\{\eta(x,t)\} - A/h_0|}{A/h_0},
\end{equation*}
where $\max\limits_x\{\eta(x,t)\}$ denotes the discrete maximum of the numerical solution and $A/h_0 = 0.1$ is the exact solitary wave amplitude. In both computations a uniform grid of 512 nodes is used. Here, again, we notice a better performance of the WN solver compared to that of the CP solution. This specific experiment shows that the WN model is a better model compared to the CP solution when nonlinear effects must be included for the study of tsunami generation and propagation.

\subsection{The July 17, 2006 tsunami generation simulation}

The main purpose of this study is to present a novel methodology for tsunami generation problems. This approach is illustrated on the example of the July 17, 2006 Java tsunami since this event is not completely understood yet and there is an available finite fault solution for the presumed generating underwater earthquake.

In this section we show a practical application of the WN method for water waves generated by a moving bottom. Namely, we exploit the bottom motion \eqref{eq:zeta} constructed in Section \ref{sec:dyndisp}. The corresponding hydrodynamic problem is solved by the three methods discussed above: the linearized water wave problem (CP), BBM-BBM system and the novel WN model.

The solution given by the WN model and the exact solutions to the linearized Euler equations \eqref{eq:lin1} -- \eqref{eq:lin2} are computed on a uniform grid of $512 \times 512$ points. The time step $\Delta t$ is chosen adaptively according to the RK(4,5) method proposed in \cite{Dormand1980}. The BBM-BBM system is solved on a triangular unstructured grid of $86276$ elements. The time integration is performed with the classical RK2 scheme \cite{Hairer2009} with time step $\Delta t = 0.5$ s.

\begin{figure}%
\centering
\subfigure[$t=20$ $s$]{\includegraphics[scale=0.35]{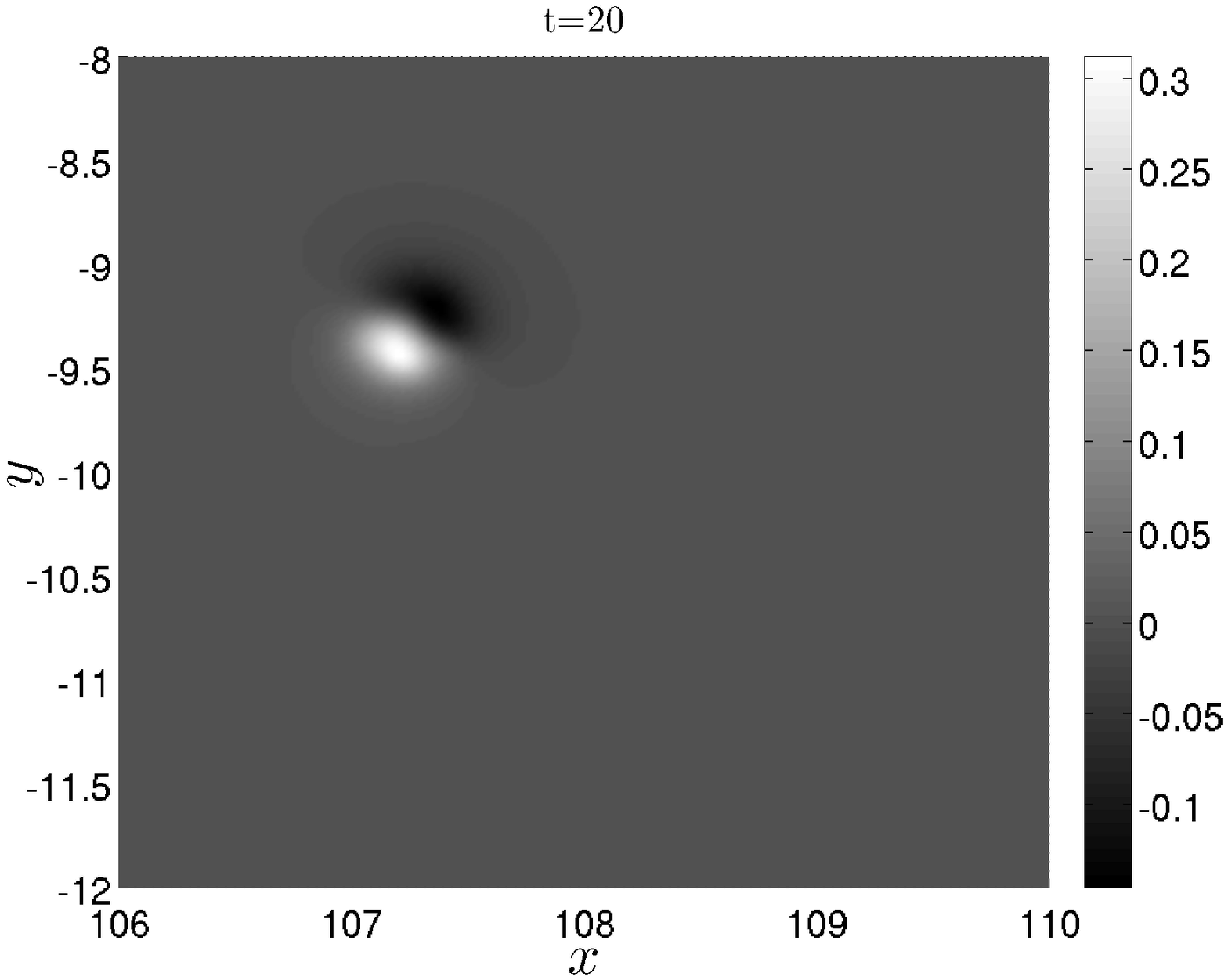}}
\subfigure[$t=50$ $s$]{\includegraphics[scale=0.35]{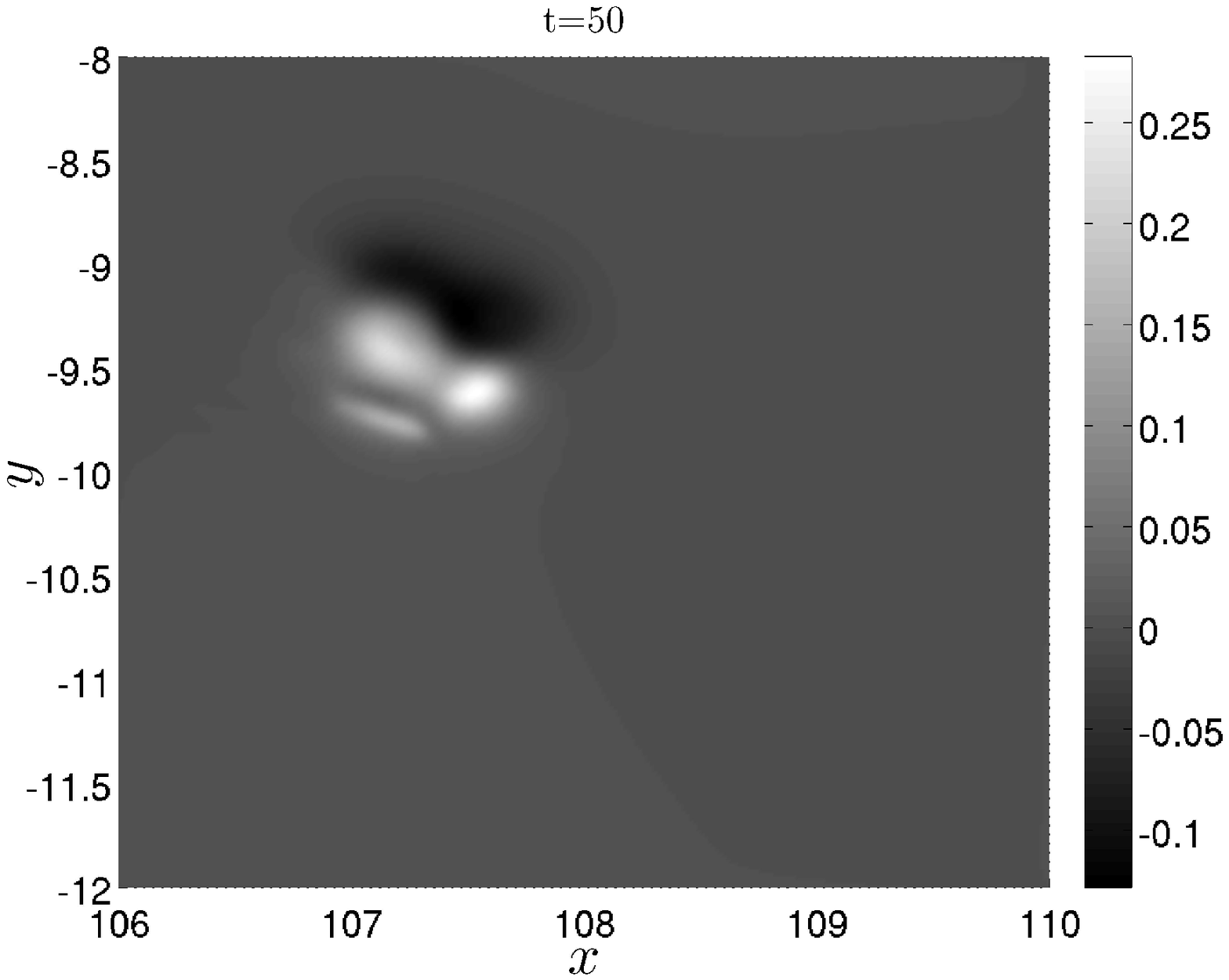}}
\subfigure[$t=80$ $s$]{\includegraphics[scale=0.35]{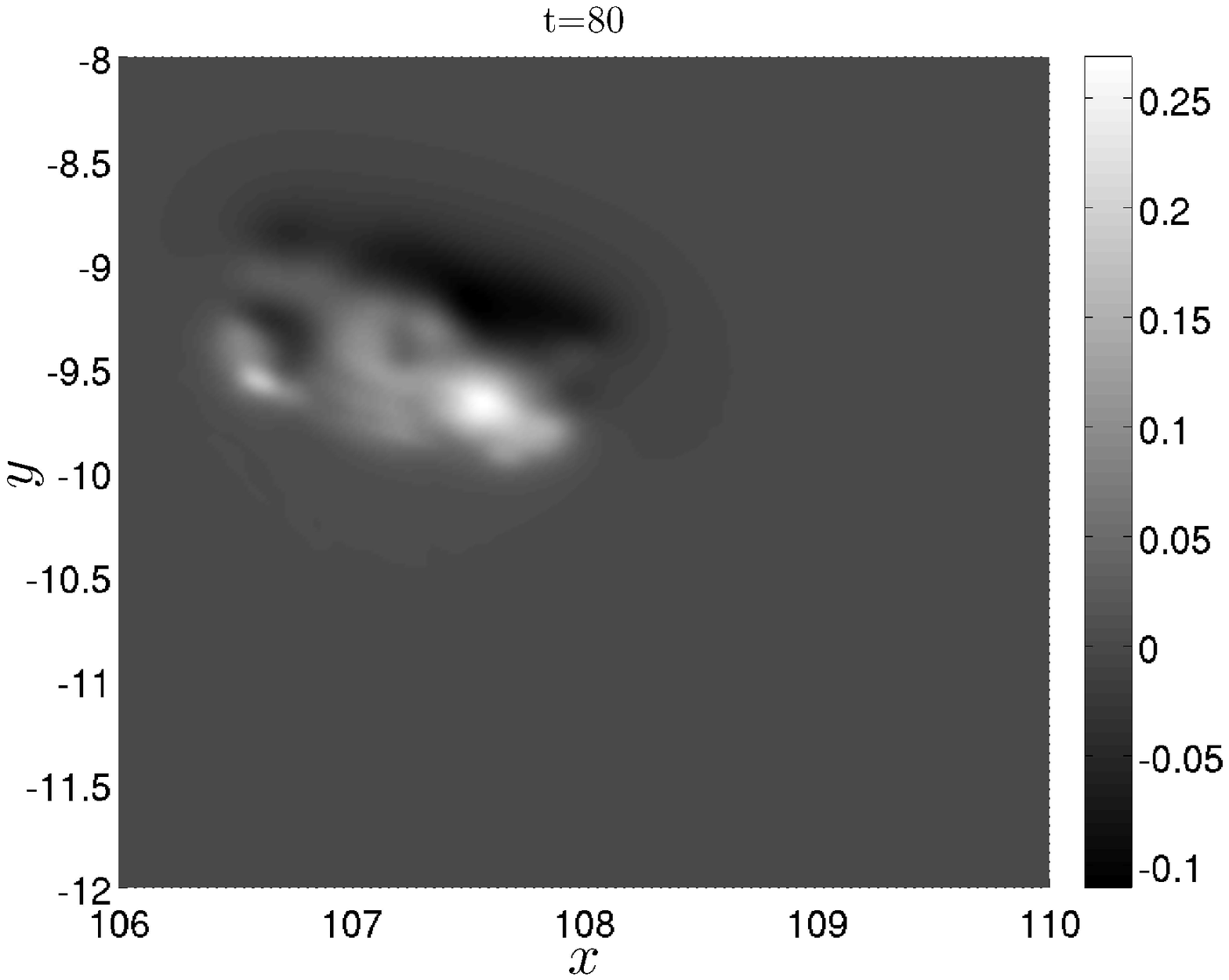}}
\subfigure[$t=140$ $s$]{\includegraphics[scale=0.35]{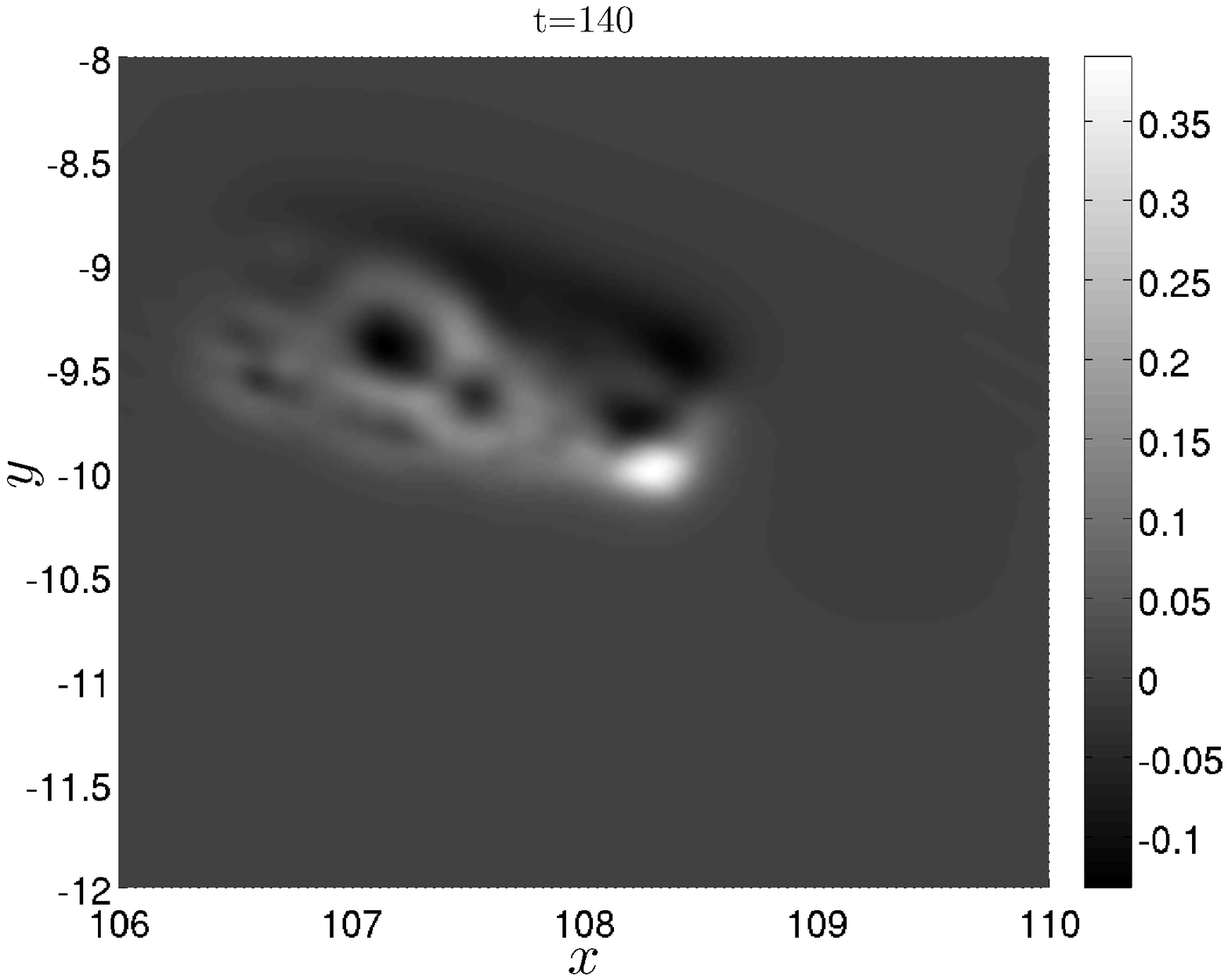}}
\subfigure[$t=200$ $s$]{\includegraphics[scale=0.35]{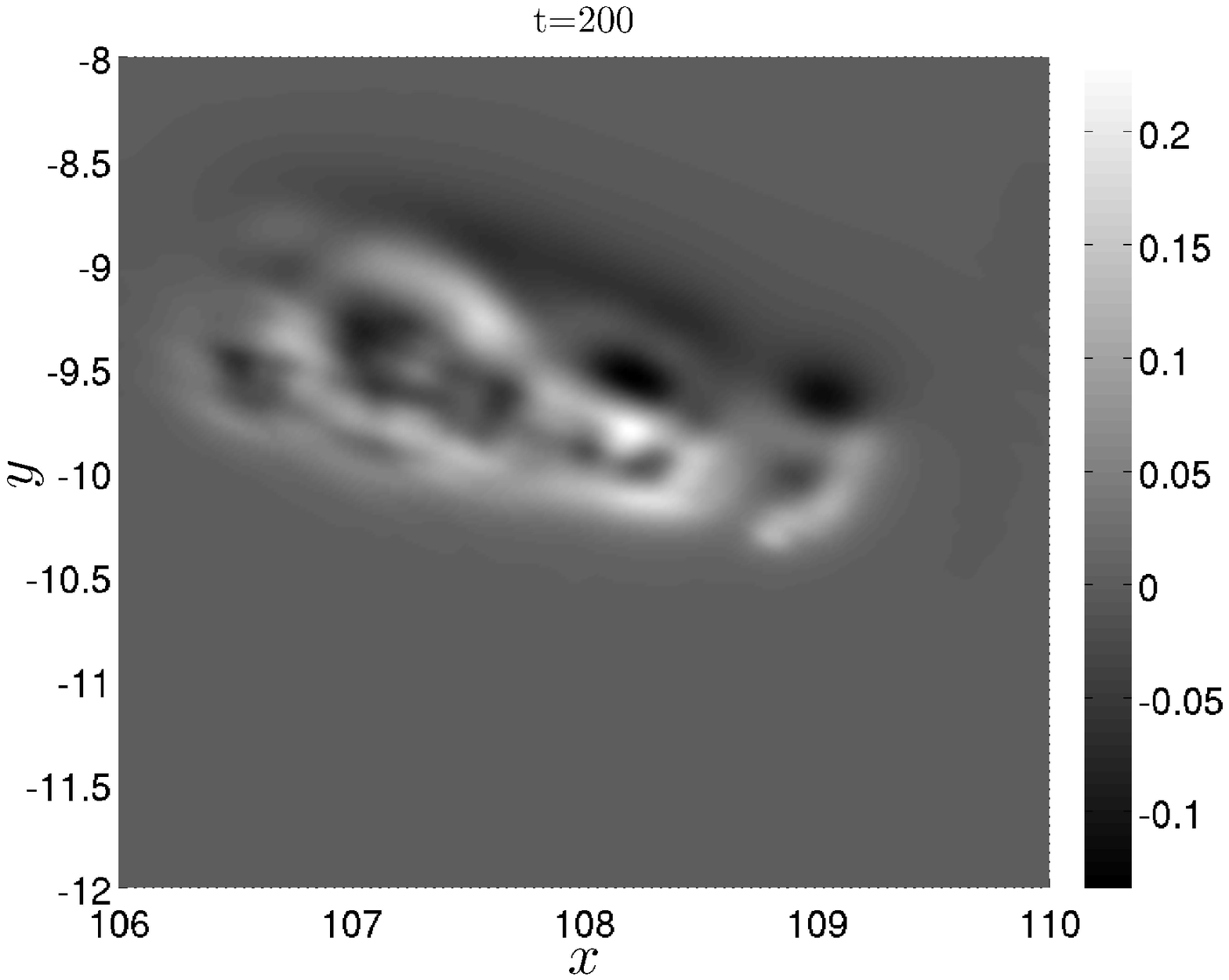}}
\subfigure[$t=250$ $s$]{\includegraphics[scale=0.35]{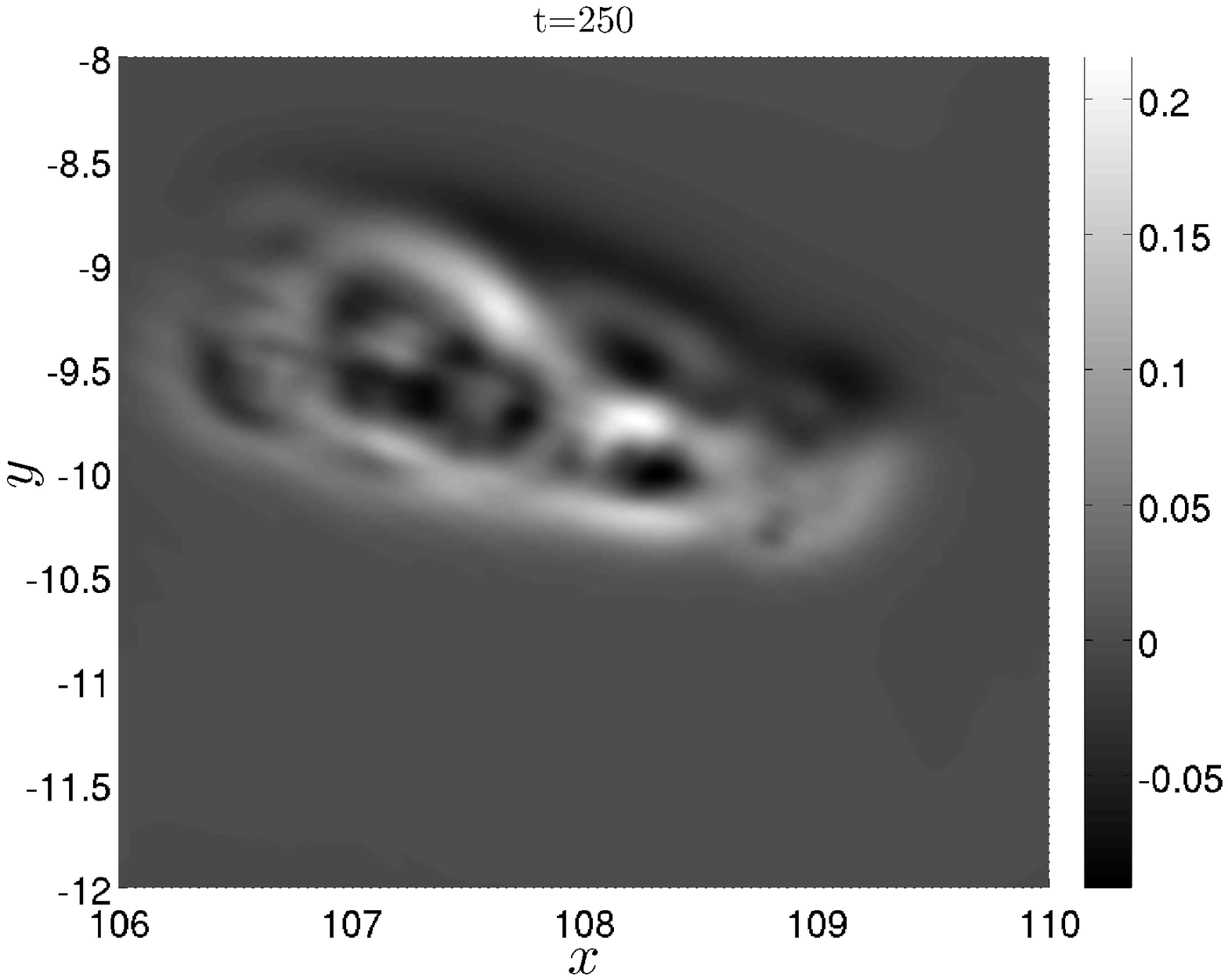}}
\caption{Snapshots of the free surface elevation computed with the weakly nonlinear (WN) model. Water waves are generated by dynamic co-seismic bottom displacements \eqref{eq:zeta} reconstructed using the corresponding finite fault solution \cite{Ji2006}.}%
\label{fig:soleuler}%
\end{figure}

Several snapshots of the free surface elevation computed with the WN model are shown in Figures \ref{fig:soleuler} (a) -- (f). Analogous contour plots of the solutions of the CP and BBM-BBM models are almost identical and differences cannot be observed within graphical accuracy. Therefore, they are not presented here. The parameters of the bottom motion, bathymetry and computational domain geometry were explained in Section \ref{sec:displ}.

\begin{figure}%
\centering
\subfigure[Gauge at $(107.2^\circ, -9.388^\circ)$]%
{\includegraphics[scale=0.25]{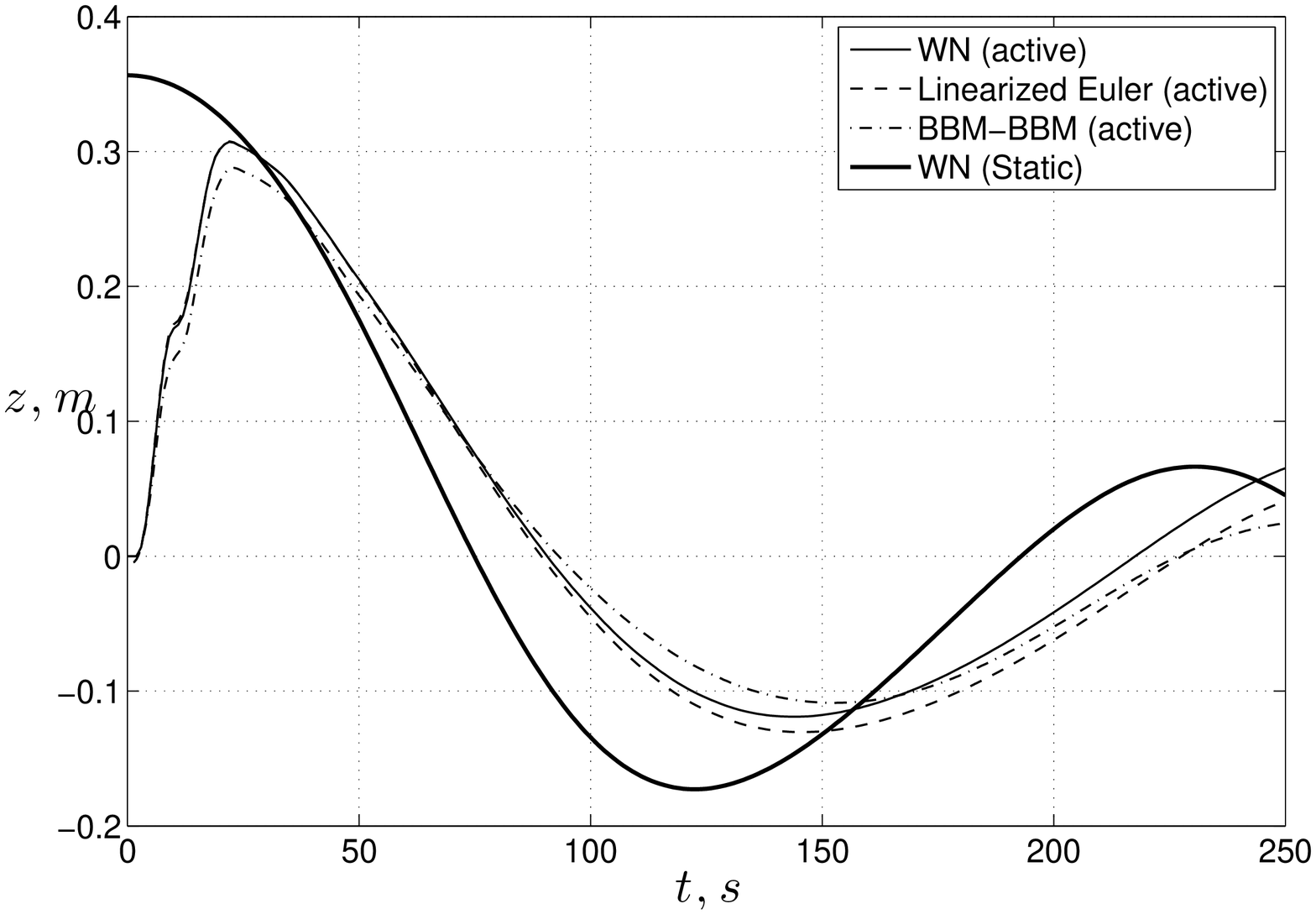}}
\subfigure[Gauge at $(107.4^\circ, -9.205^\circ)$]%
{\includegraphics[scale=0.25]{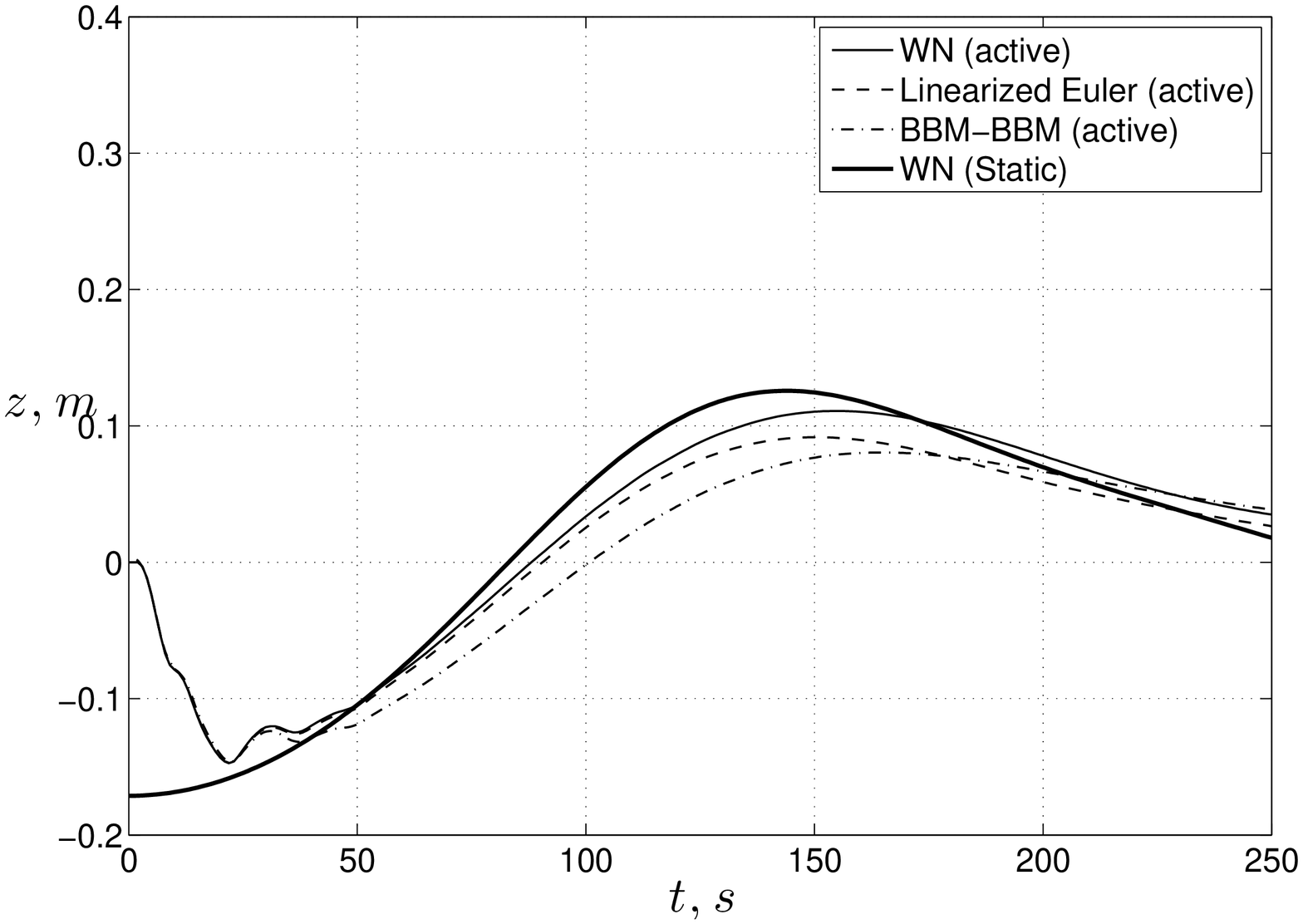}}
\subfigure[Gauge at $(107.6^\circ, -9.648^\circ)$]%
{\includegraphics[scale=0.25]{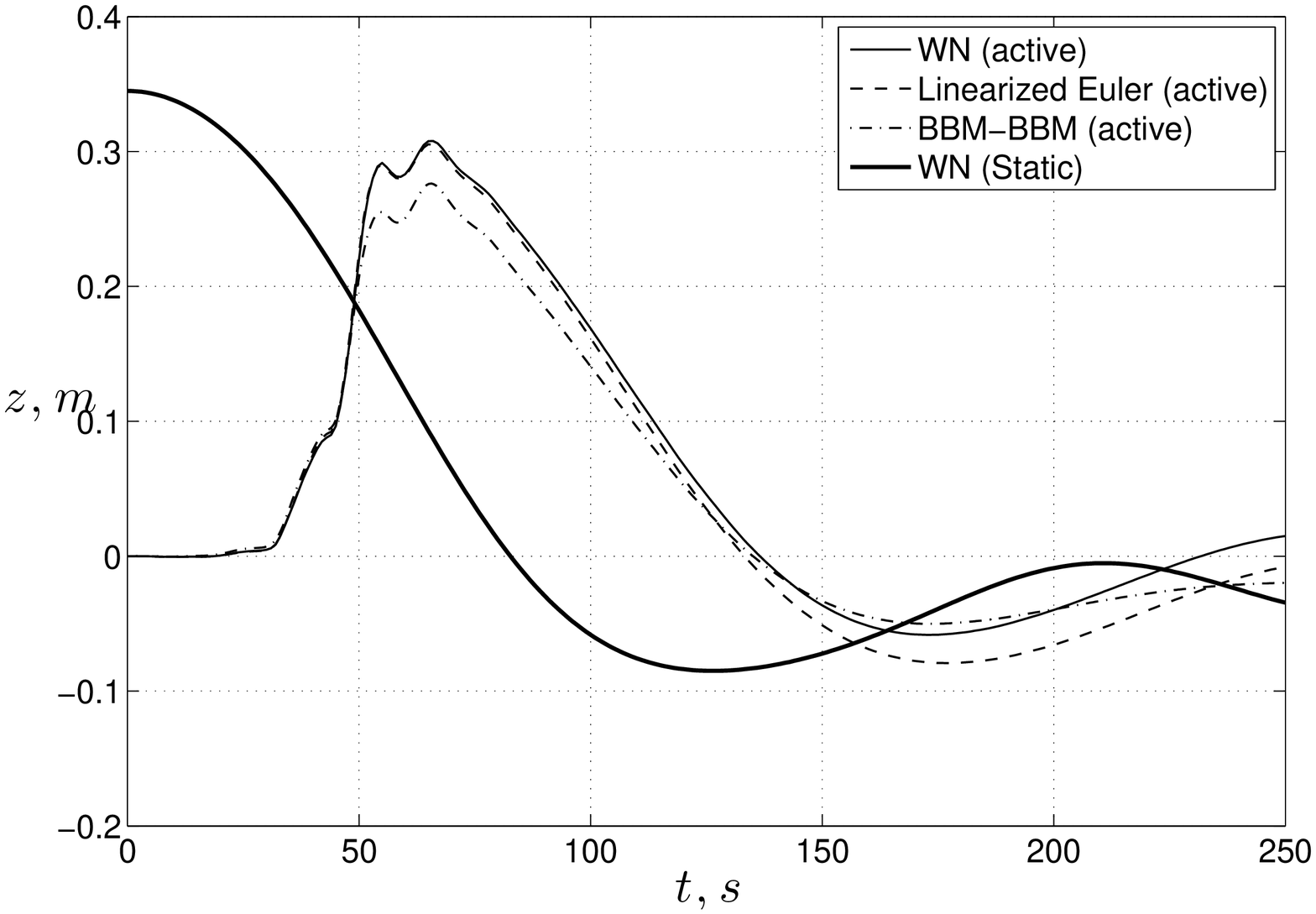}}
\subfigure[Gauge at $(107.7^\circ, -9.411^\circ)$]%
{\includegraphics[scale=0.25]{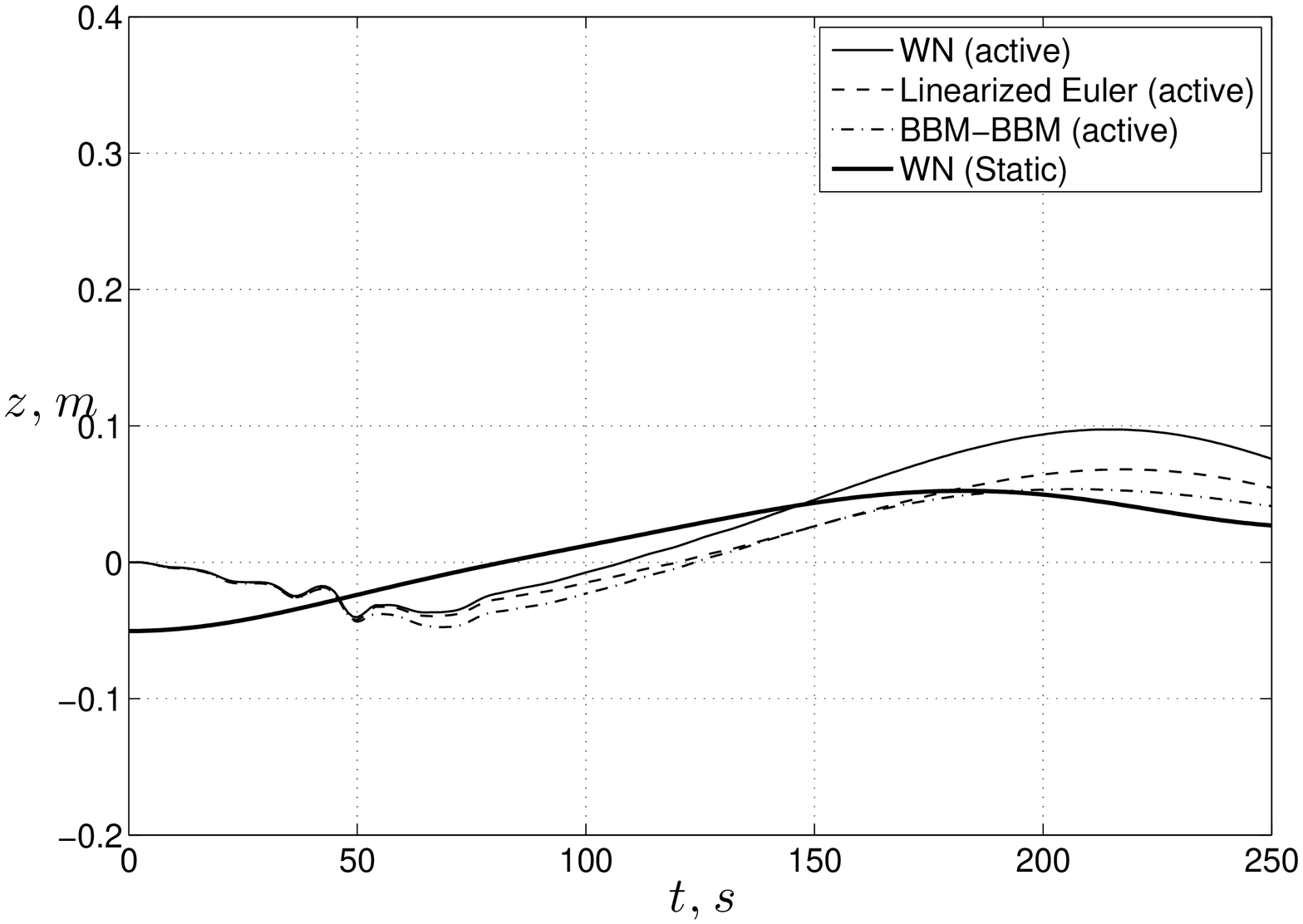}}
\subfigure[Gauge at $(108.3^\circ, -10.02^\circ)$]%
{\includegraphics[scale=0.25]{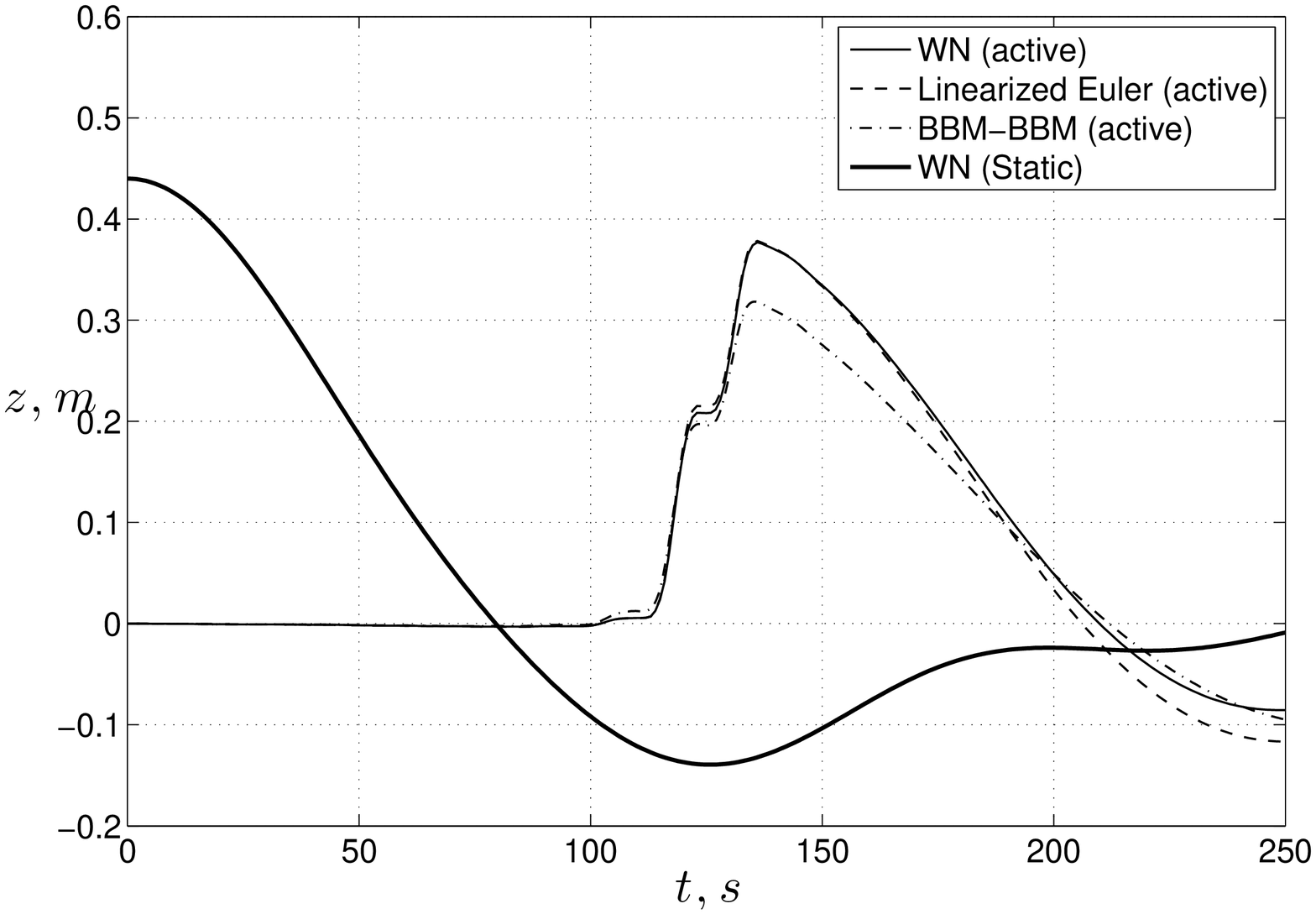}}
\subfigure[Gauge at $(108.2^\circ, -9.75^\circ)$]%
{\includegraphics[scale=0.25]{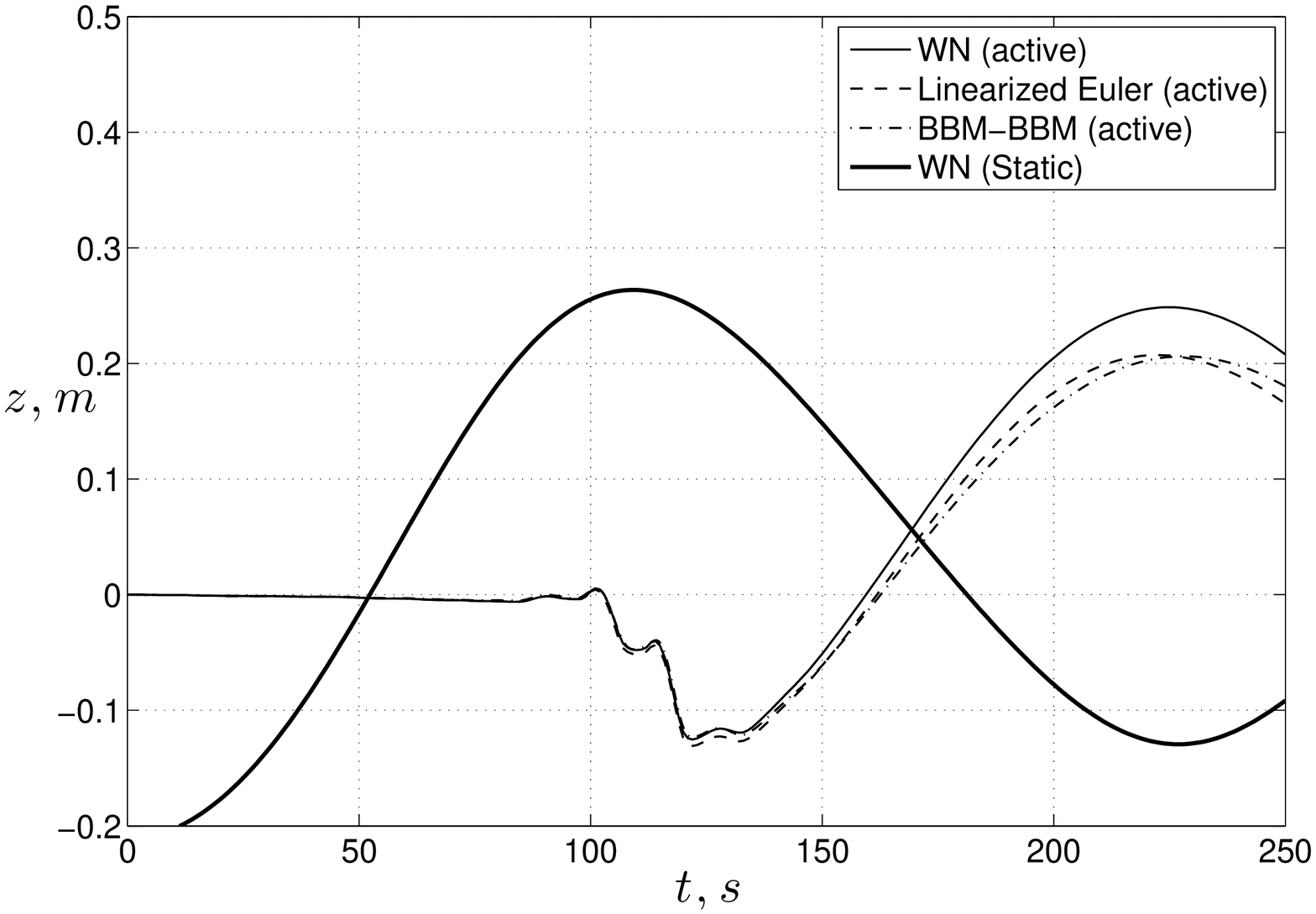}}
\subfigure[Gauge at $(108^\circ, -10.5^\circ)$]%
{\includegraphics[scale=0.25]{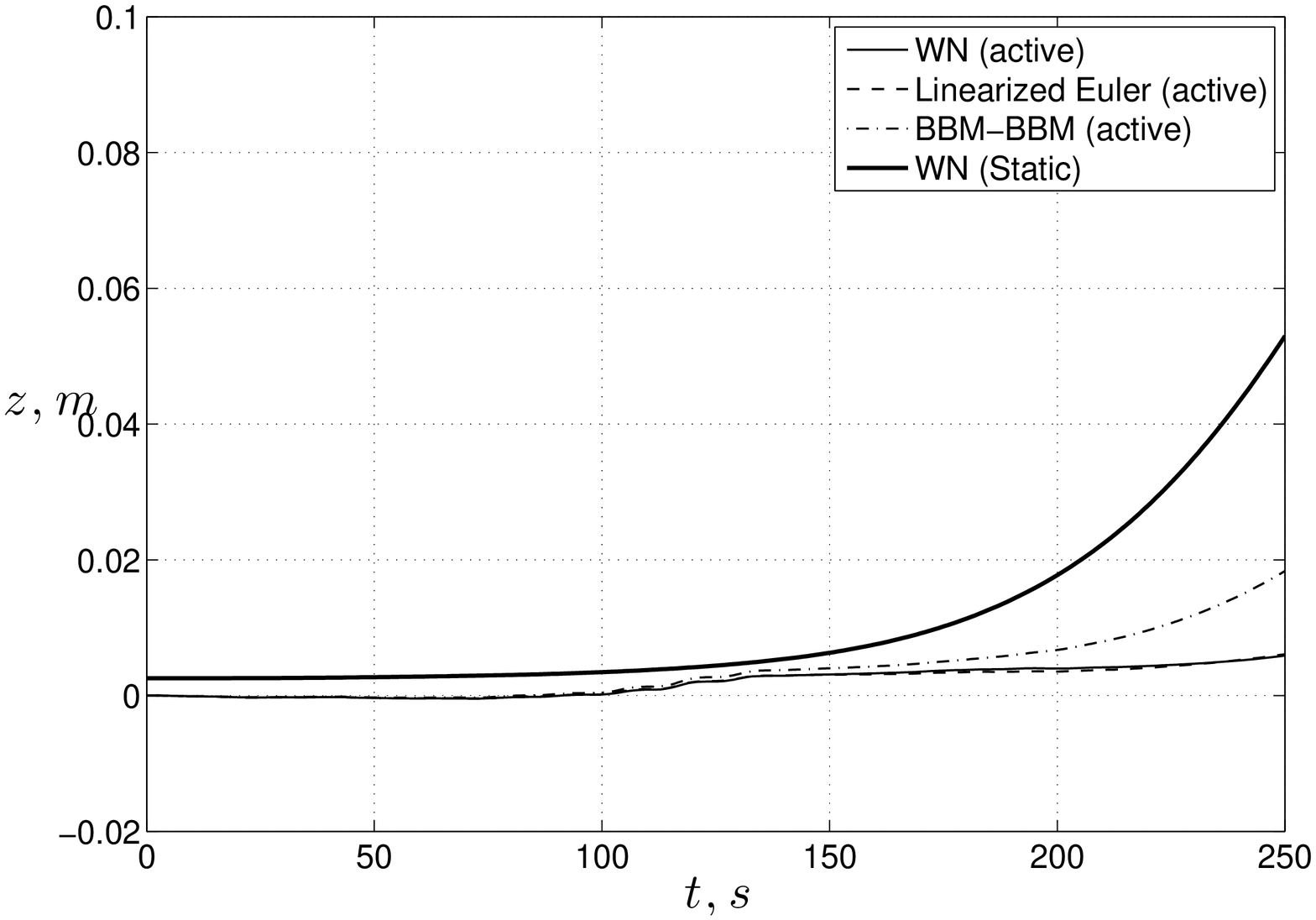}}
\subfigure[Gauge at $(108^\circ, -9^\circ)$]%
{\includegraphics[scale=0.25]{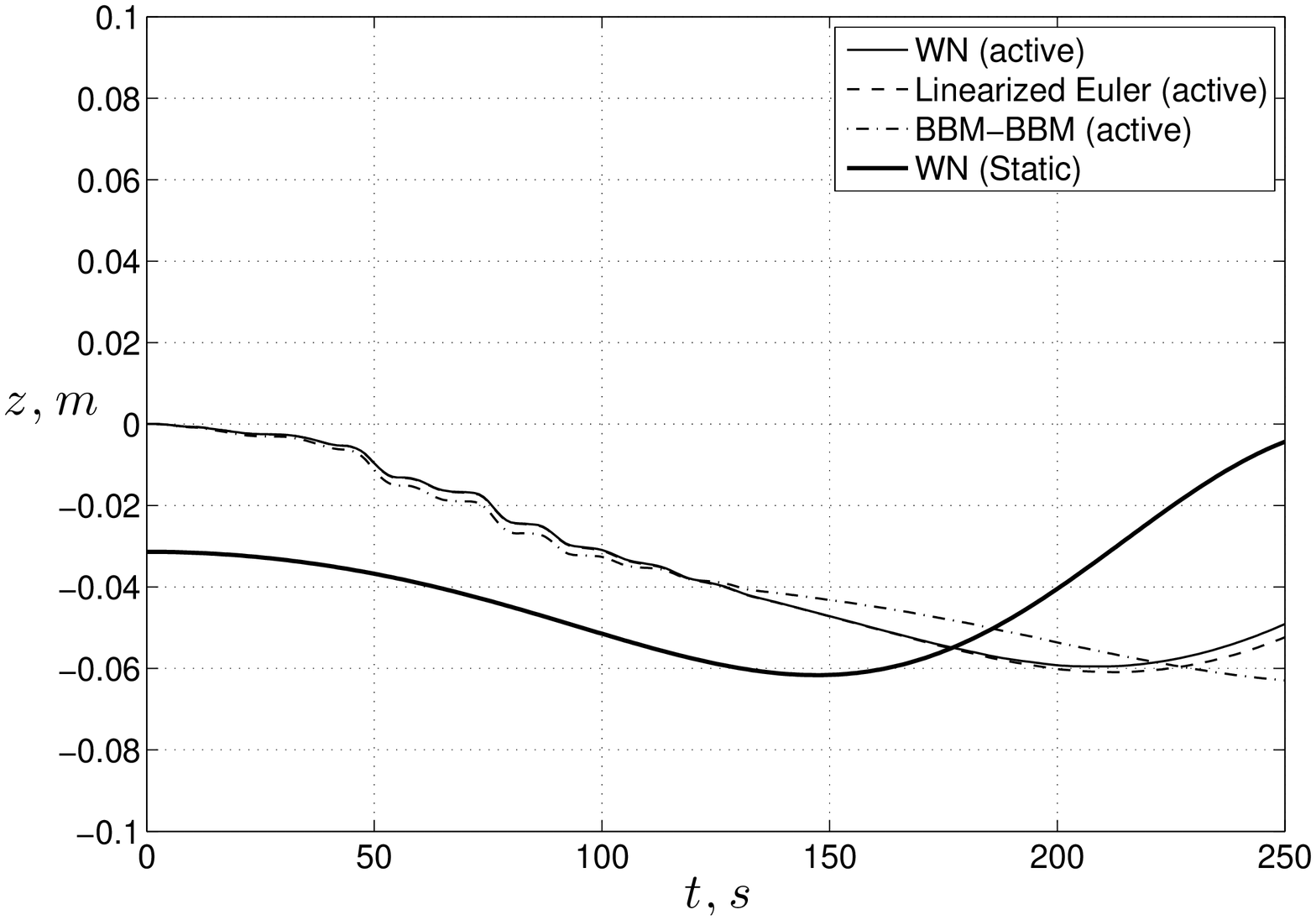}}
\caption{Free surface elevation computed numerically with four models at eight wave gauges located approximately at the local extrema of the static bottom displacement. The elevation (vertical axis) is expressed in meters, while the time (horizontal axis) is in seconds. The models 1, 2 and 4 use the dynamic finite-fault rupture (weakly nonlinear model, linearized Euler equations, BBM-BBM model). The third model uses the static approach (weakly nonlinear model).}%
\label{fig:gauges}%
\end{figure}

In this computation, we see a complex process of simultaneous wave evolution together with rupture propagation during approximately 210 s. Namely, the free surface deformed by the rupture of the first subfaults evolves while the rupture continues to propagate along the fault. This kind of fluid/moving bottom interaction cannot be described in the static generation framework, cf. Figure \ref{fig:gauges}.

\begin{figure}%
\centering
\includegraphics[scale=0.59]{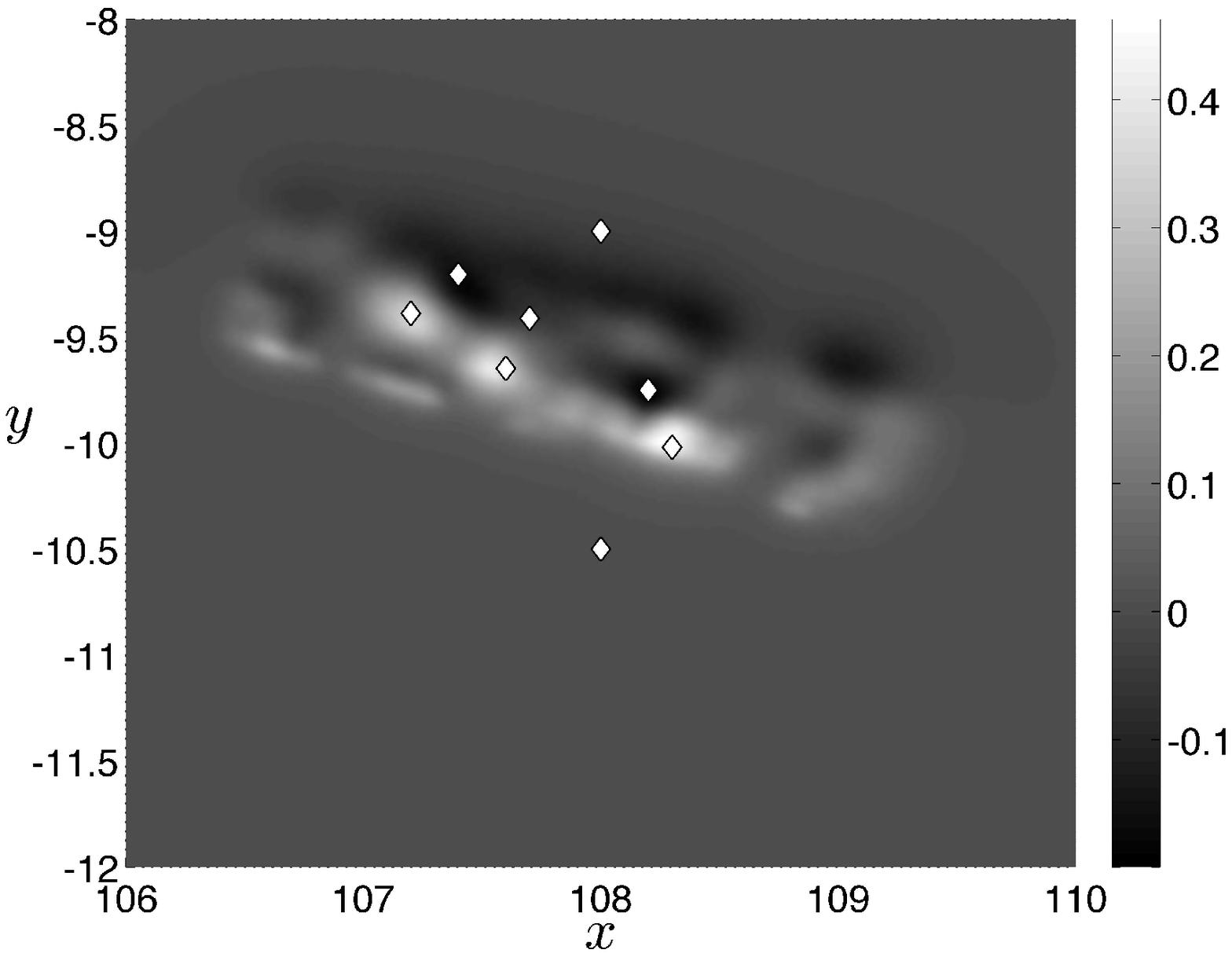}
\caption{Location of the eight numerical wave gauges (indicated by the symbol $\diamond$) superposed with the static co-seismic bottom displacement.}%
\label{fig:gaugespos}%
\end{figure}

In order to compare the three models described above we put eight numerical wave gauges at the following locations: six close to the source ((a) $(107.2^\circ$, $-9.388^\circ)$, (b) $(107.4^\circ$, $-9.205^\circ)$, (c) $(107.6^\circ$, $-9.648^\circ)$, (d) $(107.7^\circ$, $-9.411^\circ)$, (e) $(108.3^\circ$, $-10.02^\circ)$, (f) $(108.2^\circ$, $-9.75^\circ)$) and two further away from the source area ((g) $(108^\circ$, $-10.5^\circ)$, (h) $(108^\circ$, $-9^\circ)$). The locations of the wave gauges are represented by the symbol $\diamond$ on Figure \ref{fig:gaugespos} along with the static sea bed displacement. 

The eight wave gauge records are presented in Figures \ref{fig:gauges} (a)--(h). In order to show the importance of the dynamics of the rupture process, the records obtained from the static approach are also included. The overall agreement among the three dynamic models appears to be satisfactory. We underline that the CP solution is very close to the other solutions despite the fact that the bathymetric features are neglected. We also note that the specific BBM-BBM type system underestimates by a small amount the maximum wave amplitude compared to the WN model. Further numerical tests showed some sensitivity of the BBM-BBM solution to the bottom motion scenario \cite{Dutykh2006}. Namely, we can report, for example, that the exponential scenario led to a slighty larger wave amplitude compared to the other models. As expected, the static approach exhibits differences both in the shape and in the arrival time of the waves. Further away from the source area, the CP solution continues to be accurate. This is due to the fact that nonlinearity is not important during the propagation stage of such small amplitude waves. 

\section{Conclusions and perspectives}\label{sec:concl}

In the present work we considered an important issue in the modeling of tsunami generation. Namely, a new method for the construction of dynamic co-seismic sea bed displacements was proposed. This method basically relies on two main ingredients:
\begin{itemize}
  \item the finite fault solution \cite{Bassin2000, Ji2002} gives the slip distribution along the fault
  \item dynamic sea bed deformation scenarios \cite{Hammack, ddk, Dutykh2006} allow us to take into account available information of the rupture dynamics
\end{itemize}
To our knowledge, this reconstruction of the bottom motion is new. All developments presented in this paper are illustrated on the example of the July, 17 2006 Java event.

Along with the bottom motion construction, we discussed three models to solve approximately the corresponding hydrodynamic problem and compute the induced free surface motions. The July 17, 2006 tsunami generation case was computed with three different models and a comparison was performed. We obtained a surprisingly good agreement between the CP solution and the solutions of the other two models. Recall that in the latter the bottom is assumed to be flat. Discrepancies will appear later in time since the bathymetry plays a crucial role in the tsunami propagation.

Taking into account the simplicity and the relatively good accuracy of the new WN approximation to the full water wave problem with time dependent variable bottom, we suggest its use for the computation of the initial stages ($\approx 300$ $s$) of the life of a tsunami. The propagation and runup can be computed afterwards by other sophisticated tools \cite{Titov1997, Imamura2006, noaa_report, Dutykh2009a}, some of them being already integrated into tsunami warning systems \cite{Titov2005, Weinstein2008}.

However we point out that extreme runup values measured after the July, 17 Java 2006 event \cite{Fritz2007} deserve additional numerical studies.

\appendix

\section{Finite fault parameters}\label{app:i}

\begin{center}
\small
\begin{longtable}{c|c|c|c|c}
\caption{Subfault parameters given by the finite fault inversion \cite{Ji2006}.}
\label{tab:subfaults} \\

 Latitude, ${}^\circ$ & Longitude, ${}^\circ$ & Depth, $km$ & Slip, $cm$ & Rake, ${}^\circ$ \\ \hline\hline
\endfirsthead
 
 Latitude, ${}^\circ$ & Longitude, ${}^\circ$ & Depth, $km$ & Slip, $cm$ & Rake, ${}^\circ$ \\ \hline
 \endhead
 
 \hline \multicolumn{5}{c}{{Continued on the next page}} \\ \hline
 \endfoot

 \hline\hline
 \endlastfoot
 
 -10.33298 & 109.17112 & 6.81260 & 5.01844 & 121.65860 \\
 -10.28919 & 109.04183 & 6.81260 & 4.31652 & 80.93857 \\
 -10.24541 & 108.91254 & 6.81260 & 48.94745 & 85.43047 \\
 -10.20162 & 108.78325 & 6.81260 & 3.60585 & 101.68500 \\
 -10.15784 & 108.65396 & 6.81260 & 0.86479 & 67.04596 \\
 -10.11405 & 108.52467 & 6.81260 & 0.96921 & 99.45411 \\
 -10.07027 & 108.39538 & 6.81260 & 0.62447 & 71.54340 \\
 -10.02648 & 108.26609 & 6.81260 & 0.02449 & 99.44887 \\
 -9.98270 & 108.13680 & 6.81260 & 2.71502 & 119.63240 \\
 -9.93891 & 108.00751 & 6.81260 & 0.57000 & 114.25760 \\
 -9.89513 & 107.87822 & 6.81260 & 14.54725 & 112.71920 \\
 -9.85134 & 107.74893 & 6.81260 & 31.66312 & 107.26750 \\
 -9.80756 & 107.61964 & 6.81260 & 2.74176 & 85.79224 \\
 -9.76377 & 107.49035 & 6.81260 & 3.35868 & 78.97166 \\
 -9.71999 & 107.36105 & 6.81260 & 67.95367 & 64.89334 \\
 -9.67620 & 107.23177 & 6.81260 & 62.33453 & 65.43832 \\
 -9.63242 & 107.10248 & 6.81260 & 35.33318 & 66.90181 \\
 -9.58863 & 106.97318 & 6.81260 & 1.75233 & 101.93900 \\
 -9.54485 & 106.84389 & 6.81260 & 40.63542 & 81.77631 \\
 -9.50106 & 106.71461 & 6.81260 & 84.20831 & 68.95723 \\
 -9.45728 & 106.58531 & 6.81260 & 25.12981 & 66.62241 \\
 -10.24093 & 109.20313 & 8.78887 &  0.68254 & 88.79068 \\
 -10.19714 & 109.07384 & 8.78887 & 30.70282 & 97.90491 \\
 -10.15336 & 108.94455 & 8.78887 & 76.07102 & 99.93182 \\
 -10.10957 & 108.81525 & 8.78887 & 0.56201 & 79.59160 \\
 -10.06579 & 108.68597 & 8.78887 & 0.95023 & 114.32920 \\
 -10.02201 & 108.55668 & 8.78887 & 64.78191 & 121.81120 \\
 -9.97822 & 108.42738 & 8.78887 & 81.31910 & 105.21240 \\
 -9.93443 & 108.29810 & 8.78887 & 137.60680 & 121.72020 \\
 -9.89065 & 108.16881 & 8.78887 & 85.81732 & 88.13734 \\
 -9.84686 & 108.03951 & 8.78887 & 30.61069 & 80.38488 \\
 -9.80308 & 107.91022 & 8.78887 & 60.08308 & 113.75000 \\
 -9.75929 & 107.78094 & 8.78887 & 46.98381 & 96.25403 \\
 -9.71551 & 107.65164 & 8.78887 & 21.69421 & 80.82516 \\
 -9.67173 & 107.52235 & 8.78887 & 11.01957 & 112.63110 \\
 -9.62794 & 107.39307 & 8.78887 & 27.85978 & 75.88463 \\
 -9.58416 & 107.26377 & 8.78887 & 5.96505 & 77.66200 \\
 -9.54037 & 107.13448 & 8.78887 & 3.85634 & 83.57522 \\
 -9.49658 & 107.00520 & 8.78887 & 3.23158 & 113.73070 \\
 -9.45280 & 106.87590 & 8.78887 & 29.89915 & 116.10890 \\
 -9.40902 & 106.74661 & 8.78887 & 65.25044 & 72.60931 \\
 -9.36523 & 106.61732 & 8.78887 & 19.62932 & 65.99193 \\
 -10.14888 & 109.23514 & 10.76514 & 20.60663 & 124.43320 \\
 -10.10510 & 109.10584 & 10.76514 & 69.91051 & 122.64720 \\
 -10.06131 & 108.97655 & 10.76514 & 63.10052 & 99.23547 \\
 -10.01753 & 108.84727 & 10.76514 & 0.63700 & 74.09311 \\
 -9.97374 & 108.71797 & 10.76514 & 1.02761 & 117.53560 \\
 -9.92996 & 108.58868 & 10.76514 & 85.54328 & 123.64950 \\
 -9.88617 & 108.45940 & 10.76514 & 167.18620 & 104.56840 \\
 -9.84239 & 108.33010 & 10.76514 & 202.60880 & 122.12460 \\
 -9.79860 & 108.20081 & 10.76514 & 144.76970 & 81.50333 \\
 -9.75482 & 108.07152 & 10.76514 & 53.97212 & 72.84430 \\
 -9.71103 & 107.94223 & 10.76514 & 79.21021 & 98.66053 \\
 -9.66725 & 107.81294 & 10.76514 & 82.95619 & 80.81979 \\
 -9.62346 & 107.68365 & 10.76514 & 119.13390 & 74.36982 \\
 -9.57968 & 107.55436 & 10.76514 & 95.90159 & 116.24710 \\
 -9.53589 & 107.42507 & 10.76514 & 36.94965 & 102.32060 \\
 -9.49211 & 107.29578 & 10.76514 & 0.28681 & 81.49704 \\
 -9.44832 & 107.16649 & 10.76514 & 8.06018 & 98.40840 \\
 -9.40454 & 107.03720 & 10.76514 & 3.02927 & 116.89820 \\
 -9.36075 & 106.90791 & 10.76514 & 10.73559 & 74.60908 \\
 -9.31697 & 106.77862 & 10.76514 & 57.94233 & 75.39254 \\
 -9.27318 & 106.64933 & 10.76514 & 60.97223 & 64.77096 \\
 -10.05684 & 109.26714 & 12.74141 & 21.97392 & 121.10740 \\
 -10.01305 & 109.13785 & 12.74141 & 74.47045 & 119.75060 \\
 -9.96927 & 109.00856 & 12.74141 & 17.25334 & 124.09410 \\
 -9.92548 & 108.87927 & 12.74141 & 14.38904 & 87.41515 \\
 -9.88170 & 108.74998 & 12.74141 & 3.03040 & 106.36440 \\
 -9.83791 & 108.62069 & 12.74141 & 8.97587 & 101.53580 \\
 -9.79413 & 108.49140 & 12.74141 & 114.85160 & 115.94270 \\
 -9.75034 & 108.36211 & 12.74141 & 91.90382 & 115.95240 \\
 -9.70656 & 108.23282 & 12.74141 & 64.72478 & 100.08050 \\
 -9.66277 & 108.10353 & 12.74141 & 17.30368 & 123.06770 \\
 -9.61899 & 107.97424 & 12.74141 & 57.09099 & 68.20686 \\
 -9.57520 & 107.84495 & 12.74141 & 64.81193 & 79.84035 \\
 -9.53142 & 107.71566 & 12.74141 & 131.04410 & 76.45924 \\
 -9.48763 & 107.58636 & 12.74141 & 112.11020 & 99.51801 \\
 -9.44385 & 107.45708 & 12.74141 & 60.23628 & 97.77266 \\
 -9.40006 & 107.32778 & 12.74141 & 126.96870 & 80.27277 \\
 -9.35628 & 107.19849 & 12.74141 & 63.39000 & 65.00801 \\
 -9.31249 & 107.06921 & 12.74141 & 0.52621 & 94.79313 \\
 -9.26871 & 106.93991 & 12.74141 & 1.52171 & 66.78681 \\
 -9.22492 & 106.81062 & 12.74141 & 10.96743 & 81.94861 \\
 -9.18114 & 106.68134 & 12.74141 & 2.38062 & 123.04830 \\
 -9.96479 & 109.29915 & 14.71768 & 22.40949 & 123.90350 \\
 -9.92100 & 109.16986 & 14.71768 & 48.62879 & 115.45630 \\
 -9.87722 & 109.04057 & 14.71768 & 5.99559 & 83.81007 \\
 -9.83343 & 108.91128 & 14.71768 & 7.22945 & 123.80940 \\
 -9.78965 & 108.78199 & 14.71768 & 0.10031 & 93.40998 \\
 -9.74586 & 108.65269 & 14.71768 & 0.36991 & 69.37087 \\
 -9.70208 & 108.52341 & 14.71768 & 104.18760 & 123.83230 \\
 -9.65829 & 108.39411 & 14.71768 & 46.12533 & 95.97049 \\
 -9.61451 & 108.26482 & 14.71768 & 0.28679 & 89.56866 \\
 -9.57072 & 108.13554 & 14.71768 & 2.06597 & 80.14312 \\
 -9.52694 & 108.00624 & 14.71768 & 30.55070 & 66.23147 \\
 -9.48315 & 107.87695 & 14.71768 & 73.72994 & 87.91253 \\
 -9.43937 & 107.74767 & 14.71768 & 112.90700 & 92.28181 \\
 -9.39558 & 107.61837 & 14.71768 & 74.73608 & 86.51558 \\
 -9.35180 & 107.48908 & 14.71768 & 121.73820 & 64.68654 \\
 -9.30801 & 107.35979 & 14.71768 & 231.20940 & 65.50779 \\
 -9.26423 & 107.23050 & 14.71768 & 96.55727 & 87.01543 \\
 -9.22044 & 107.10121 & 14.71768 & 28.29534 & 122.55670 \\
 -9.17666 & 106.97192 & 14.71768 & 0.84110 & 70.21989 \\
 -9.13287 & 106.84263 & 14.71768 & 7.99213 & 87.51706 \\
 -9.08909 & 106.71334 & 14.71768 & 1.33281 & 96.33266 \\
 -9.87274 & 109.33115 & 16.69394 & 43.31154 & 121.79150 \\
 -9.82896 & 109.20187 & 16.69394 & 87.17052 & 124.49750 \\
 -9.78517 & 109.07257 & 16.69394 & 61.47630 & 87.10537 \\
 -9.74139 & 108.94328 & 16.69394 & 31.53286 & 70.58137 \\
 -9.69760 & 108.81400 & 16.69394 & 0.70628 & 65.17896 \\
 -9.65382 & 108.68470 & 16.69394 & 5.74160 & 87.70702 \\
 -9.61003 & 108.55541 & 16.69394 & 93.47714 & 107.32000 \\
 -9.56625 & 108.42612 & 16.69394 & 93.55753 & 85.39201 \\
 -9.52246 & 108.29683 & 16.69394 & 47.25525 & 74.24297 \\
 -9.47868 & 108.16754 & 16.69394 & 24.65230 & 124.20110 \\
 -9.43489 & 108.03825 & 16.69394 & 35.63115 & 71.78733 \\
 -9.39111 & 107.90896 & 16.69394 & 25.11757 & 75.27779 \\
 -9.34732 & 107.77967 & 16.69394 & 68.15302 & 107.42980 \\
 -9.30354 & 107.65038 & 16.69394 & 24.66007 & 112.77880 \\
 -9.25975 & 107.52109 & 16.69394 & 0.50688 & 79.86887 \\
 -9.21597 & 107.39180 & 16.69394 & 119.92850 & 75.03103 \\
 -9.17218 & 107.26250 & 16.69394 & 77.08335 & 110.83160 \\
 -9.12840 & 107.13322 & 16.69394 & 31.65430 & 123.83060 \\
 -9.08461 & 107.00393 & 16.69394 & 11.42768 & 66.47282 \\
 -9.04083 & 106.87463 & 16.69394 & 33.80650 & 115.65650 \\
 -8.99704 & 106.74535 & 16.69394 & 39.47481 & 65.15574 \\
 -9.78069 & 109.36316 & 18.67021 & 35.42621 & 111.95830 \\
 -9.73691 & 109.23387 & 18.67021 & 103.05030 & 124.62650 \\
 -9.69312 & 109.10458 & 18.67021 & 101.38220 & 122.70620 \\
 -9.64934 & 108.97529 & 18.67021 & 76.76701 & 68.20042 \\
 -9.60556 & 108.84600 & 18.67021 & 10.71945 & 77.79713 \\
 -9.56177 & 108.71671 & 18.67021 & 1.32449 & 100.72950 \\
 -9.51799 & 108.58742 & 18.67021 & 37.46857& 124.59330 \\
 -9.47420 & 108.45813 & 18.67021 & 118.99580 & 100.38000 \\
 -9.43042 & 108.32883 & 18.67021 & 79.62616 & 91.56905 \\
 -9.38663 & 108.19955 & 18.67021 & 97.61735 & 109.86430 \\
 -9.34285 & 108.07026 & 18.67021 & 87.67753 & 87.57239 \\
 -9.29906 & 107.94096 & 18.67021 & 15.14859 & 64.75201 \\
 -9.25528 & 107.81168 & 18.67021 & 82.60960 & 71.66805 \\
 -9.21149 & 107.68239 & 18.67021 & 66.06397 & 98.55843 \\
 -9.16771 & 107.55309 & 18.67021 & 0.43085 & 67.81042 \\
 -9.12392 & 107.42381 & 18.67021 & 35.30429 & 124.04570 \\
 -9.08014 & 107.29452 & 18.67021 & 59.17323 & 124.55130 \\
 -9.03635 & 107.16522 & 18.67021 & 15.23214 & 66.82615 \\
 -8.99257 & 107.03593 & 18.67021 & 28.10358 & 76.08198 \\
 -8.94878 & 106.90664 & 18.67021 & 48.09923 & 124.24450 \\
 -8.90500 & 106.77735 & 18.67021 & 42.38682 & 124.42850 \\
\end{longtable}
\end{center}

\section{Zakharov's formulation of the water wave problem}\label{app:ii}

In this appendix we recast the governing equations (\ref{eq:laplace}) -- (\ref{eq:bottomkin}) of the water wave problem in a more compact and mathematically more convenient form \cite{Zakharov1968, Craig1993}.

Using the definition of the normal velocity (\ref{eq:normalv}), it is straightforward to rewrite the kinematic free surface condition (\ref{eq:kinematic}):
\begin{equation*}
  \dt\eta - \D_\eta (\phis) = 0,
\end{equation*}
where $\phis$ is the trace of the velocity potential at the free surface \eqref{eq:trace}.

The time derivative and the horizontal gradient of the velocity potential trace on the free surface can be computed:
\begin{equation}\label{eq:tphis}
  \dt\phis = \dt\phi + \dt\eta \left.\dz\phi\right|_{z=\eta} = \dt\phi + \D_\eta(\phis)\left.\dz\phi\right|_{z=\eta},
\end{equation}
and similarly one can compute the horizontal gradient:
\begin{equation}\label{eq:gradphis}
  \grad\phis = \left.\grad\phi\right|_{z=\eta} + \grad\eta \left.\dz\phi\right|_{z=\eta}.
\end{equation}
In order to close the system, we have to express all derivatives of the potential $\phi$ computed at the free surface, in terms of $\phis$, $\eta$ and $\D_\eta(\phis)$.

From the definition of the normal velocity (\ref{eq:normalv}) and the D2N operator one readily obtains:
\begin{equation}\label{eq:phizeta}
  \left.\grad\phi\right|_{z=\eta}\cdot\grad\eta = 
  \left.\dz\phi\right|_{z=\eta} - \D_\eta(\phis).
\end{equation}
Substituting the last identity into (\ref{eq:gradphis}) multiplied by $\grad\eta$, leads to the following expression:
\begin{equation}\label{eq:phiz}
  \left.\dz\phi\right|_{z=\eta} = 
  \frac{\D_\eta(\phis) + \grad\phis\cdot\grad\eta}{1 + |\grad\eta|^2}.
\end{equation}
Now we have all elements to find the horizontal derivatives of the velocity potential:
\begin{equation}\label{eq:gradphi}
  \left.\grad\phi\right|_{z=\eta} = \grad\phis - \grad\eta \left.\dz\phi\right|_{z=\eta}
  = \frac{(1 + |\grad\eta|^2)\grad\phis - \D_\eta(\phis)\grad\eta 
  - (\grad\phis\cdot\grad\eta)\grad\eta} {1 + |\grad\eta|^2}.
\end{equation}

In order to rewrite Bernoulli condition (\ref{eq:bernoulli}) in new variables, we make the following observation (using (\ref{eq:gradphis}) and (\ref{eq:phizeta})):
\begin{multline*}
  \half|\grad\phi|^2 + \half(\dz\phi)^2 = \half\grad\phi\cdot\grad\phi +
  \half\dz\phi\;\dz\phi = \\ = \half\grad\phi\cdot(\grad\phis - \dz\phi\grad\eta) + 
  \half\dz\phi(\D_\eta(\phis) + \grad\phi\cdot\grad\eta) = 
  \half\grad\phi\cdot\grad\phis + \half\D_\eta(\phis)\dz\phi, \quad z=\eta.
\end{multline*}
Taking into account this observation and expression (\ref{eq:tphis}) for the time derivative of $\phis$, the dynamic condition takes this equivalent form:
\begin{equation*}
  \dt\phis + g\eta + \half\grad\phi\cdot\grad\phis - \half\D_\eta(\phis)\dz\phi = 0,
  \quad z = \eta.
\end{equation*}
After substituting expressions (\ref{eq:phiz}), (\ref{eq:gradphi}) into the last equation and summarizing all the developments made above, we get the following set of dynamic equations equivalent to the complete water wave problem (\ref{eq:laplace}) -- (\ref{eq:bottomkin}):
\begin{equation*}
 \begin{array}{rl}
  \dt\eta - \D_\eta(\phis) &= 0, \\
  \dt\phis + \half|\grad\phis|^2 + g\eta
  - \frac{1}{2(1+|\grad\eta|^2)}\bigl[\D_\eta(\phis) + 
   \grad\phis\cdot\grad\eta \bigr]^2 &= 0.
 \end{array}
\end{equation*}

\section{Relations between elastic constants}\label{app:iii}

In the classical elasticity theory, coefficients in Lam\'{e} equations (governing the displacements field in an elastic solid), can be expressed in terms of various sets of physical parameters \cite{love, sokol}. The purpose of this Appendix is to recall some relations between them.

Lam\'e coefficients $\lambda$ and $\mu$ can be defined in terms of the Young's modulus $E$ (having the dimension of the pressure $[Pa]$) and Poisson's ratio $\nu$ (dimensionless coefficient $0<\nu<1/2$):
\begin{equation*}
  \lambda = \frac{E\nu}{(1+\nu)(1-2\nu)}, \quad
  \mu = \frac{E}{2(1+\nu)},
\end{equation*}
and inversely:
\begin{equation*}
  E = \frac{(3\lambda + 2\mu)\mu}{\lambda + \mu}, \quad
  \nu = \frac{\lambda}{2(\lambda + \mu)}.
\end{equation*}
The celerities of $P$ and $S$ waves have the following expressions in terms of Lam\'e coefficients:
\begin{equation*}
  c_p = \sqrt{\frac{\lambda + 2\mu}{\rho}}, \quad 
  c_s = \sqrt{\frac{\mu}{\rho}},
\end{equation*}
where $\rho$ is the density of elastic medium. These relations yield
\begin{equation*}
  \mu = \rho c_s^2, \quad 
  \lambda = \rho c_p^2 - 2\mu.
\end{equation*}

\section*{Acknowledgement}
D.~Dutykh acknowledges the support from French Agence Nationale de la Recherche, project MathOcean (Grant ANR-08-BLAN-0301-01). F.~Dias and D.~Dutykh acknowledge the support of the Ulysses Program of the French Ministry of Foreign Affairs under the project 23725ZA.

Special thanks go to Professor Costas Synolakis whose work on tsunami waves has been the source of our inspiration. Finally we would like to thank Professor Didier Clamond for very helpful discussions on the numerical simulation of water waves.

\bibliography{biblio}
\bibliographystyle{alpha}

\end{document}